\documentclass[twocolumn]{aastex63}
\usepackage{epsfig}
\usepackage{natbib}
\usepackage{amsmath}
\usepackage{hyperref}
\usepackage{comment}

\usepackage{xcolor}
\definecolor{forestgreen}{rgb}{0.13, 0.75, 0.13}

\newcommand{\blue}{\color{black}}
\newcommand{\red}{\color{black}}

\newcommand{\p}{^\prime}

\newcommand{\zindex}{z}

\begin{document}

\title{The Pantheon+ Analysis: SuperCal-Fragilistic Cross Calibration, Retrained SALT2 Light Curve Model, and Calibration Systematic Uncertainty}

\author{Dillon Brout}
\affil{Center for Astrophysics, Harvard \& Smithsonian, 60 Garden Street, Cambridge, MA 02138, USA} 
\affil{NASA Einstein Fellow}
\email{dillon.brout@cfa.harvard.edu}

\author{Georgie Taylor}
\affil{Research School of Astronomy and Astrophysics, Australian National University, Canberra, Australia}

\author{Dan Scolnic}
\affil{Department of Physics, Duke University, Durham, NC, 27708, USA.}

\author{Charlotte M. Wood}
\affil{Department of Physics, University of Notre Dame, Notre Dame, IN 46556, USA}

\author{Benjamin M. Rose}
\affil{Department of Physics, Duke University, Durham, NC, 27708, USA.}

\author{Maria Vincenzi}
\affil{Department of Physics, Duke University, Durham, NC, 27708, USA.}

\author{Arianna Dwomoh}
\affil{Department of Physics, Duke University, Durham, NC 27708, USA}

\author{Christopher Lidman}
\affil{The Research School of Astronomy and Astrophysics, Australian National University, ACT 2601, Australia}
\affil{Centre for Gravitational Astrophysics, College of Science, Australian National University, ACT 2601, Australia}

\author{Adam Riess}
\affil{Department of Physics and Astronomy, Johns Hopkins University, Baltimore, MD 21218, USA}

\author{Noor Ali}
\affil{Umeå University, 901 87, Umeå, Sweden}

\author{Helen Qu}
\affil{Department of Physics and Astronomy, University of Pennsylvania, Philadelphia, PA 19104, USA}

\author{Mi Dai}
\affil{Department of Physics and Astronomy, Johns Hopkins University, Baltimore, MD 21218, USA}



\begin{abstract}
We present a re-calibration of the photometric systems in the Pantheon+ sample of Type Ia supernovae (SNe Ia) including those in the SH0ES distance-ladder measurement of H$_0$. We utilize the large and uniform sky coverage of the public Pan-STARRS stellar photometry catalog to cross-calibrate against tertiary standards released by individual SN Ia surveys. The most significant updates over the `SuperCal'  cross-calibration used for the previous Pantheon and SH0ES analyses are: 1) expansion of the number of photometric systems (now 25) and filters (now 105), 2) solving for all filter offsets in all systems simultaneously to produce a calibration uncertainty covariance matrix for cosmological-model constraints, and 3) accounting for the change in the fundamental flux calibration of the HST CALSPEC standards from previous versions on the order of $1.5\%$ over a $\Delta \lambda$ of 4000~\AA. We retrain the SALT2 model and find that our new model coupled with the new calibration of the light-curves themselves causes a net distance modulus change ($d\mu/dz$) of 0.04 mag over the redshift range $0<z<1$. We introduce a new formalism to determine the systematic impact on cosmological inference by propagating the covariance in fitted calibration offsets through retraining simultaneously with light-curve fitting and find a total calibration uncertainty impact of $\sigma_w=0.013$; roughly half the size of the sample statistical uncertainty.  Similarly, we find the systematic SN calibration contribution to the SH0ES H$_0$ uncertainty is less than 0.2~km/s/Mpc, suggesting that SN Ia calibration cannot resolve the current level of the `Hubble Tension'.

\end{abstract}




\section{Introduction}

Type Ia supernovae (SNe Ia) are a critical tool to measure the expansion history of the universe.  They are particularly useful for measuring the recent cosmic acceleration and thus the equation-of-state of dark energy $w$ and the current expansion rate $H_0$.  In cosmological analyses with SNe Ia, calibration of the photometric system is typically one of the largest systematics in the error budget \citep{Betoule2014,Scolnic18,Jones19}.  The calibration errors are survey and filter dependent, which typically manifest in redshift-dependent changes in distance because 1) different surveys cover different redshift ranges and 2) SNe redshift into different observer frame wavelengths. In order to maximize statistical leverage and minimize the impact of calibration errors when constraining cosmological parameters, recent analyses have combined SNe from multiple different photometric systems.  In this paper, we perform an up-to-date recalibration of photometric systems used in the Pantheon+ sample \citep{scolnic22} and cosmological analysis (Brout et al in prep), and propagate these changes through light-curve model training/fitting and to cosmological inference.

Photometric calibration is required in two critically important components of SN Ia cosmological analyses. {\blue First, the calibration of different SN light-curves in a `training library’ must be accounted for in order to build the spectral time-series model that will be used to fit light-curve parameters of a larger photometric sample.  Second, the calibration of light-curves in the full sample must be accounted for to apply the model and fit for light-curve parameters and recover distance estimates. Importantly, the calibration of light-curves used in the training library should be self-consistent with the calibration of the light-curves used in the larger sample, but historically this has not always been the case.  Recent cosmological analyses \citep{Scolnic18,Brout18b} of SNe Ia used the SALT2 model from \cite{Betoule2014} (hereafter B14) because the SALT2 model has not been available for retraining.  Therefore, all current analyses have not benefited from improved calibration since B14.  However, because of recent work to update and make available SALT2 retraining code \citep{Taylor21,Kenworthy21}, it is now possible to retrain the light-curve model with the same calibration used in fitting.}

The Pantheon+ sample (\citealt{scolnic22}, hereafter S21) compiles data from 25 different photometric systems. To perform the cross calibration, we follow the framework described in \cite{Scolnic2015} (hereafter SuperCal), which used Pan-STARRS (PS1, \citealp{Chambers16}) photometry of tertiary standard stars to recalibrate each photometric system, as PS1 covers 3$\pi$ of the sky and has sufficient overlap with each survey.  An update to the SuperCal process was presented in \cite{Currie20} (hereafter, `Excalibur'), which followed the same premise, but implemented a number of changes including transferring every system onto its natural system and and fitting for filter transformations. In this work we implement a number of the recommendations from Excalibur, but retain the simplicity of SuperCal of no additional transformations to natural system, no measurements across the field-of-view of each camera, and we focus on simultaneous fitting photometric magnitude offsets (zeropoints) rather than fitting filter transformations. 

This paper is a companion to a suite of papers ({\red e.g. S21, \citealt{carr2021pantheon,peterson21,sasha21,popovic21,binningissinning,itsdust,dhawan20H0}}) leading up to the Pantheon+ cosmological analysis (Brout et al in prep), which contains the measurements of cosmic acceleration, dark energy, and dark matter, and SH0ES distance ladder Hubble constant analysis \citep{riess22}. This work focuses solely on the cross-calibration of the samples and associated systematic uncertainty. In Section 2 we describe the data sample and overview the suite of calibration systems used by various SN analyses.  In Section 3 we describe the re-calibration process and compare with past results.  In Section 4, we discuss the light-curve model retraining.  In Section 5, we propagate these changes towards the impact on cosmological inference and produce a new systematic error budget for calibration.  In Sections 6 \& 7, we present our discussions and conclusions.

\section{Overview of Synthetic and Observed Data}

\subsection{Individual Survey Calibration}

{\blue Different photometric analyses have calibrated their surveys through a variety of paths.  Most recent analyses of SNe Ia have tied their absolute calibration of their photometric systems to the AB system \citep{oke83,fukijita96}, which has mostly replaced the historical use of absolute calibration to the Vega flux standard \citep{vega}. Many older SN samples have alternatively used tertiary standards from \cite{landolt92} or \cite{smith02}, which themselves can be externally tied to a fundamental calibration like the AB system.}

In the AB system, broadband magnitudes are defined as
\begin{equation}
m_{\rm AB} = 2.5 \times {\rm log}_{10}\frac{\int h\nu^{-1}~p(\nu)~f_\nu ~d\nu}{\int h\nu^{-1}~p(\nu)~3631 {\rm J_y}~d\nu}
\label{eq:one}
\end{equation}
where $p(\nu)$ is the transmission function of a given filter, $f_\nu$ is the flux per unit frequency from an object in J$_{\rm y}$ and 3631 J$_{\rm y}$ corresponds to the flux of a monochromatic 0th magnitude object.

For systems calibrated to the Landolt system, linear photometric transformations for each filter are used to bring standard magnitudes of tertiary stars to the natural-system magnitudes following

\begin{equation}
   m_{{\rm nat},f} = m_f  - \epsilon_f \times C
\end{equation}
where $\epsilon_f$ is the color coefficient for filter $f$ and $C$ is the color from the standard magnitude (i.e. (B - V )). 

For both AB and Landolt/Vega calibration methods, specific passband magnitude offsets are computed following 
\begin{equation}
   m_{\rm AB}= m_{{\rm nat}}  - \Delta_{\rm AB}
\label{eq:three}
\end{equation}
where $\Delta_{\rm AB}$ is the offset for the particular filter to bring the system magnitude to an AB magnitude and where we have dropped the index $f$ for convenience. If a photometric system is defined on AB, then these offsets $\Delta_{\rm AB}$ are found explicitly by comparing to primary standard stars.

All systems calibrated here rely on HST CALSPEC standards \citep{bohlin96} as updated in \cite{Bohlin21}\footnote{\href{https://archive.stsci.edu/hlsps/reference-atlases/cdbs/current\_calspec/}{https://archive.stsci.edu/hlsps/reference-atlases/cdbs/current\_calspec/}}. The CALSPEC spectra data was taken with STIS and NICMOS observations and have an associated uncertainty of $\sim$1 mmag/$1000\AA$~from $3000\AA$~to $15000\AA$~\citep{bohlin14}.

   \begin{figure}[b]
        \centering 
	    \includegraphics[width=.43\textwidth]{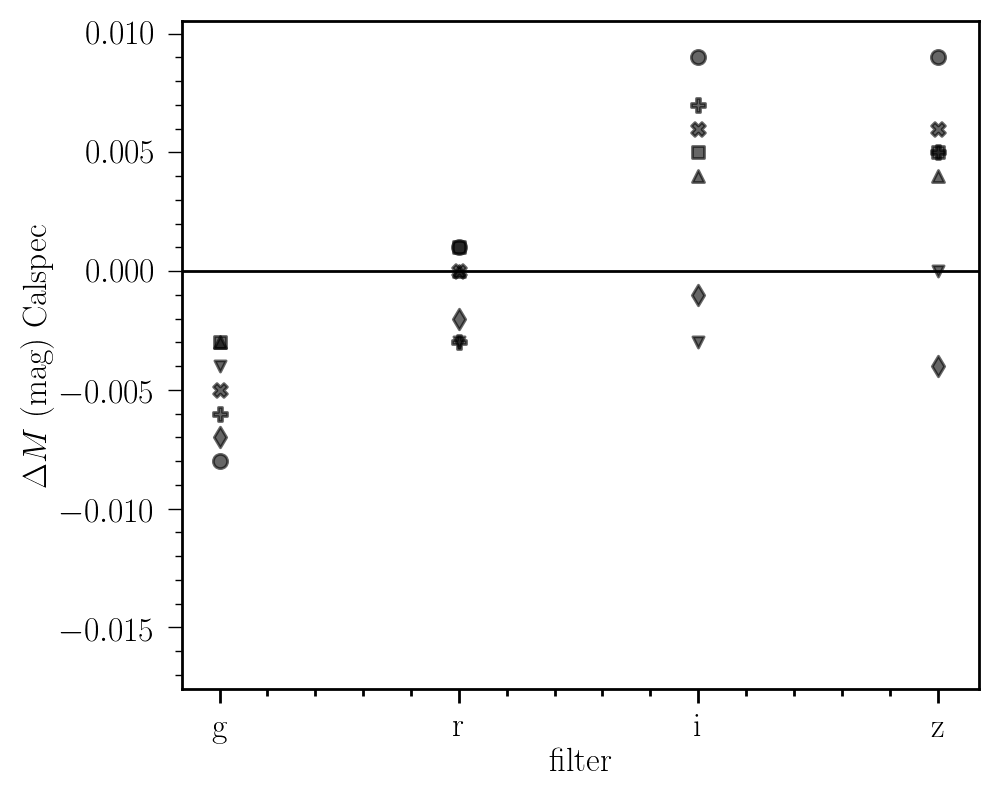} 
	    \includegraphics[width=.43\textwidth]{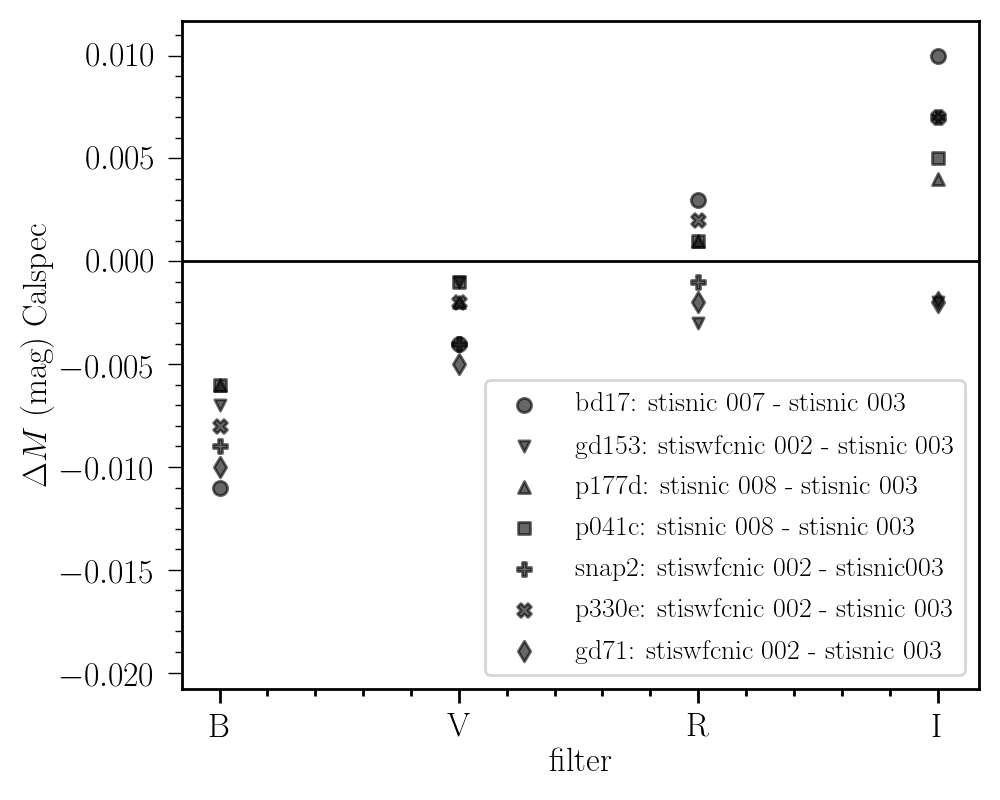} 
        \caption{Residuals between CALSPEC standard spectra integrated with standard filters are shown between the versions of CALSPEC used in \cite{Betoule2014} and those used in this work. (Top) SDSS $griz$ filters (Bottom) Bessel $BVRI$ filters.}
        \label{fig:CALSPECplot} 
    \end{figure} 
    
\subsection{Synthetic Data}
\label{sec:CALSPEC}
The spectra of flux standards are used in two ways in this analysis: 1) To establish the calibration of each system by tying observed photometry to spectrophotometry of HST CALSPEC standards and 2) use as a representative sample of stars to determine the transformation functions between different optical systems and filters thus facilitating cross calibration using observed stellar magnitudes. { In this work we do not alter the CALSPEC spectra to assess CALSPEC related systematic uncertainties. For CALSPEC systematic uncertainty and impact on cosmological analysis see \cite{pantheonpluscosmo}.}

First, while many surveys used in this work have already been calibrated to HST CALSPEC standards, the CALSPEC spectra themselves have been updated and improved {\blue significantly over the years (i.e. \citealt{bohlin96,bohlin07,bohlin14,bohlin15,bohlin19,Bohlin21}).}

{ Due to the updates in CALSPEC spectra, we homogenize the initial, independent calibrations (Eqs.~\ref{eq:one}-\ref{eq:three}) prior to performing any cross calibration,} by recalibrating the PS1, SDSS, SNLS, and DES surveys using their provided observations of CALSPEC standards and the latest spectral models of these CALSPECs as given in \cite{Bohlin21}.  {\red These offsets that homogenize PS1, SDSS, SNLS, and DES are also included in Appendix Table~\ref{tab:recal}.}

{\blue The magnitude of these changes for typical Bessel BVRI and Sloan $griz$ filters is shown in Fig.~\ref{fig:CALSPECplot}, which presents the difference in the synthetic-stellar magnitudes of CALSPEC standards between different versions of the CALSPEC spectra. In this work we will compare our results to the previous calibration \citep{Betoule2014}, so we examine the contributions due to differences arising from updates to the CALSPECs. The most recent} and improved stisnic 007/008 and stiswfcnic 002 versions of the CALSPECs results in a 1.5-2\% change from $g/B$ to $I/z$, $\sim3\times$ larger than the expected systematic uncertainty of the CALSPEC calibration of $\sim0.5\%$ over 7000 A.  {\blue These changes in the absolute calibration due to the update of the CALSPEC standards have the largest impact in our analysis when comparing to the previous \cite{Betoule2014} and SuperCal calibrations. Since this is a change in the reference, this affects the inferred zeropoint offsets of all SN samples.}

We also utilize synthetic spectral libraries to compute the transformations between photometric systems.  For this, we use two synthetic libraries: the HST CALSPEC library, as discussed above, and the NGSL\footnote{\href{https://archive.stsci.edu/prepds/stisngsl/}{https://archive.stsci.edu/prepds/stisngsl/}} library.  While the CALSPEC library has been continually updated; there is only a single release of the NGSL library \citep{ngsl}.  There are 68(370) total stars with spectra from CALSPEC(NGSL) usable for our measurements, and $74\%$ of them have $g-i<0.2$, limiting the color range that we require in order to be in the range of typical field stars. For our cross-calibration measurements, we propagate both CALSPEC and NGSL libraries and weight them equally in fitting.

\subsection{Description of Observed Data and Photometric Systems}

\begin{table}
\centering
\begin{tabular}{l|rrr}
\toprule
Sample & Filters & Recalibrated & SuperCal \\
\hline
CfA1  & $BVRI$ & N & N$^{a}$ \\
CfA2  & $BVRI$ &N & N$^{a}$ \\
HST-SCP  & F160... & N & N \\
HST-MCT  &  F160... & N & N \\
HST-GOODS  & F160... & N & N\\
\\
CfA3-Keplercam  & $BVri$ &Y & Y \\
CfA3-4Shooter  &$BVri$ & Y & Y  \\
CfA4\_p1  &$BVri$ & Y & Y  \\
CfA4\_p2  &$BVri$ & Y & Y  \\
CSP DR3  &$griBV$ & Y & Y  \\
CSP post DR3  &$griBV$ & Y & Y  \\
CSP-II  &$griBV$ & Y & Y  \\
SDSS  &  $griz$ &Y & Y \\
SNLS  &  $griz$ &Y & Y \\
PS1 SNe  & $griz$ &  Y & Y \\
\\
CNIa0.02 &$BVri$ & Y & N  \\
LOSS1  &$BVRI$ & Y & N \\
LOSS2  &$BVRI$ & Y & N \\
SOUSA &$BV$ & Y & N \\Foundation  & $griz$ & Y & N \\
DES  & $griz$ & Y & N \\

\end{tabular}
\caption{List of the data samples in Pantheon+, whether they are recalibrated in this work, and whether they were previously recalibrated in SuperCal.\phantom{AAAAAAAAAAAAAA} $^a$ SuperCal did attempt to recalibrate these surveys, but the offsets were insignificant and therefore not used in Pantheon analysis.
}
\label{tab:summary}
\end{table}

In this analysis, we have compiled photometry from $25$ different photometric systems and 105 different filters.  
There are a number of details that are critical to the reproducibility of analysis with a given photometric system.  We summarize these details in Table~\ref{tab:metainfo} and in Appendix A we fill in this table for each photometric system.

\begin{table}
\centering
\begin{tabular}{lr}
\toprule
    Question & Answer \\
\hline

Where are filters published? & Paper Reference \\
Website for the filters? & Webpage location \\
Does filter have atmospheric term?	& Yes/no \\
Where are stars published? & Paper reference \\
Website site for the stars? & Webpage location \\
What system are the stars in? & (Natural/Landolt/AB) \\
What is calibration based on? & Standard (e.g. BD17) \\
What is transformation? & Equations if applicable. \\
Need to transform stars? & Yes/no
\end{tabular}
\caption{Survey calibration metadata template. Filled out versions of this template for each survey can be found in Appendix Tables 8-21.}
\label{tab:metainfo}
\end{table}

Following SuperCal, we use the public PS1 survey photometric catalogs \citep{Chambers16} to cross-calibrate against each individual survey. The level of relative calibration across the sky reported by PS1 is $\sim5$ mmag \citep{Schlafly16}.  We use the latest, public release DR2\footnote{\href{https://catalogs.mast.stsci.edu/panstarrs/}{https://catalogs.mast.stsci.edu/panstarrs/}} all-sky coverage to perform the cross-calibration.  

\subsection{PS1 \& Foundation}
{\blue In this analysis, we utilize 3 different versions of photometry from the Pan-STARRS telescope. 
\begin{enumerate}
    \item \textbf{PS1 Public}: The aforementioned public DR2 catalog. For this sample, we use aperture magnitudes as suggested by \cite{Currie20}, as they are more robust to non-linearity than point-spread-function (PSF) photometry. {These magnitudes are used as the intermediary to perform the cross calibration between all surveys.}
    
    \item \textbf{PS1 SNe}: The set of stars that were used to calibrate the original PS1 SN Ia images \citep{Rest14,Scolnic13}.  The stellar magnitudes are based on the absolute calibration in \cite{Tonry12} and are not adjusted based on the later re-calibration for PS1 alone in SuperCal. {These magnitudes are used to determine calibration offsets for the PS1 SNe.}
    
    \item \textbf{Foundation}: The catalog used for the Foundation SN Ia sample \citep{Foley2018} taken with the PS1 telescope. {These magnitudes are used to determine calibration offsets for the Foundation SNe.}
\end{enumerate}
}

\subsection{Carnegie Supernova Project (CSP)}
The calibration of each band was rederived in \cite{krisciunis17} and \cite{krisciunas20}. This work was done after SuperCal was released, which relied on \cite{Stritzinger2010}.  It is therefore difficult to directly compare the zeropoint-corrections found in this study compared to those found in SuperCal. {\blue We compared the photometry of the same tertiary stars in the standard system presented by \cite{Stritzinger2010} and \cite{krisciunas20} and find offsets of mean (median) $g,r,i,B,V$ offsets between 2010 and the 2020 erratum of the order 1-2\% and are shown in Appendix Fig.~\ref{fig:cspudpate}. It is unclear what those offsets were due to, and we remark that we don’t see similar offsets for the SN photometry, which are all below $0.01$ mag. We therefore include additional systematic uncertainty in the calibration of CSP SNe Ia (Section~\ref{sec:sys}).}

\subsection{CfA 1 \& 2 and other Heterogenous Low-$z$ SNe Ia}
Due to the limited number of stars ($<20$ for CfA1; $<50$ for CfA2), we are unable to produce an accurate cross-calibration for these surveys.  SuperCal attempted to recalibrate these two samples, but the offsets found were not significantly deviated from 0 due to high uncertainties, so \cite{Scolnic18} did not apply the corrections from SuperCal.  We also note that two of the light curves from CfA1 (SN 1994ae and SN 1995al) were recalibrated in \cite{riess05,riess09} as they have higher importance due to their use as Cepheid-calibrators.

{\blue As described in S21, there are other heterogeneous low-$z$ datasets compiled for {\blue SH0ES} of typically O(1) SN light-curves (e.g. \citealp{Milne10,Kris2017,Stritzinger10,Tsvetkov10})}, often the stellar photometry is not provided or the number of stars is inadequate for cross-calibration. For these, because we cannot cross-calibrate, we assume a roughly 3$\times$ larger calibration uncertainty (20 mmag) than the typical reported uncertainties by other surveys, as discussed in Section 4.

\subsection{CfA3 and CfA4}
The CfA3 and CfA4 samples were taken on the F. L. Whipple Observatory 1.2m telescope's cameras: 4Shooter and Keplercam. For CfA3-4Shooter, the $UBVRI$ passbands were published in \cite{Hicken09b} and accounted for atmospheric transmission. \cite{Hicken12} stated that the first period of CfA4 (CfA4p1) and CfA3-Keplercam shared the same photometric system. The unsubmitted Cramer et al. in prep unofficially released CfA4p1 filters via private communication.  For this reason we use CfA4p1 filters for the CfA3-Keplercam stars and SNe. Neither of the CfA4p1 or CfA4p2 filters account for atmospheric transmission and this was not accounted for in SuperCal. However, for this analysis we apply a MODTRAN atmospheric transmission following \cite{stubbsandtonry}, {\red assuming typical airmass of 1.2, water vapor, and aerosols at Mt. Hopkins.}

\subsection{AB Surveys: SDSS, SNLS, \& DES}
We re-derive the AB pre-offsets (Appendix B) for SDSS and DES. For SNLS, we adopt the same changes to the AB offsets as SDSS because \cite{Betoule2014} did not provide the final observed magnitudes of the CALSPEC standards, and instead presented observations across the focal plane.  As \cite{Betoule2014} used the same version of CALSPEC photometry (v003), we employ the same shift for SNLS as SDSS, which we assume to be reliable to $\sim2$ mmag per band.  For DES, there are two calibrations, one for the SNe that were analyzed on an older FGCM star catalog for the DESSN-3YR sample and a second calibration for the upcoming DES 5 year FGCM star catalog upon which the unpublished DESSN-5YR photometric sample will be based with the intention that the DES collaboration will utilize the Fragilistic calibration solution determined here in their cosmological analyses. {\blue The offsets applied are in Appendix Table \ref{tab:recal}.}

\subsection{SOUSA}
Samples from SOUSA have not been included in previous compilation analyses like Pantheon or \cite{Betoule2014}. SOUSA has not yet had a publication that has released stars or SNe, however in private communication (with Peter Brown) we received a release of the stellar photometry performed with the same pipeline as that of the SN photometry \citep{brown14}. Because SOUSA was calibrated via the VEGA system, we use the most recent release of CALSPEC VEGA for the determination of the initial calibration (Alpha Lyr stis010). 

\subsection{LOSS}
\label{sec:loss}
There are two different data releases from the LOSS survey: \citep[LOSS1;][]{Ganeshalingam10} and \citep[LOSS2;][]{KAIT-LOSS}.  Each of the data releases includes SNe observd by the KAIT and Nickel telescopes, and the telescopes go through a series of `configurations' where the throughput for each filter is different for each configuration.  In \cite{Ganeshalingam10}, they use configurations from KAIT1-4 and Nickel1-2; in \cite{KAIT-LOSS}, they use configurations from KAIT3-4 and Nickel1-2.
Data from these systems have not been included in previous compilation analyses like Pantheon or \cite{Betoule2014}.  
 
 \cite{Ganeshalingam10} provides all tertiary standards in the Landolt system, and then provides the transformations to convert to the natural system; these are given in Appendix A and applied. For Nickel1 and Nickel2, the transformation appears to be only valid for a small color range as it produces a slope in color-color space different than what is expected. Therefore we restrict the synthetic and data color ranges ($\pm0.1$) to the median stellar color of $g-i$ of 0.75 on a typical image. \cite{KAIT-LOSS} uses the public Pan-STARRS catalogs as their tertiary standards, after they calibrate the PS1 photometry of stars to the Landolt system following transformation as given in \cite{Tonry12}.  This transformation was not specific to the filters used in \cite{KAIT-LOSS}, but rather generic $BVRI$ filters. We repeat this procedure as transcribed in \cite{KAIT-LOSS} to recreate the actual tertiary catalog they used, and use this to perform our cross-calibration.

\subsection{Complete Nearby Supernova Sample}
\label{sec:cnia}

{\blue The Complete Nearby (Redshift less than 0.02) Supernova Sample (CNIa0.02), a followup survey of ASAS-SN discovered transients, released $BVri$ filter throughputs in \cite{cnia0.02} but did not provide stars. Since they performed the same calibration procedure as \cite{KAIT-LOSS}, we utilize the stars from \cite{KAIT-LOSS} in combination with the released CNIa0.02 filter throughput curves.}

\subsection{HST SN Ia samples}
As HST observations of CALSPEC standards are themselves used to define the absolute calibration presented in \cite{Bohlin21}, we do not recalibrate the HST SNe Ia in our cross-calibration.  From past findings, we adjust the photometric zeropoints of NICMOS F105W photometry fainter by 0.068 mag and F160W fainter by 0.023 mag as the result of 3 net changes: 1) updated NICMOS zero points over original calibration in \cite{Riess07} used 2) \cite{Rubin15} check on low count rate zeropoint and 3) update to \cite{Rubin15} based on revision of WFC3 zeropoints between 2012 and 2020.  It includes the NICMOS count-rate non-linearity which is built into the recalibration of \cite{Rubin15}. Additionally, following \cite{Rubin15} we use zeropoint errors of 0.022 and 0.023 mag for F105W and F160W respectively.

\section{Calibration Solution from Simultaneous Fitting Method}

\subsection{Procedure}

{\blue For the given set of stars that overlap between PS1 and each other survey ($S$), the expected difference in magnitude between a PS1 filter (${\rm PS1}^{b1}$) and a given survey filter (${\rm Obs}_{S^{b\prime}}$) is expressed as:

\begin{equation}
\begin{aligned}
    {\rm Obs}_{{\rm PS1}^{b1}}-{\rm Obs}_{S^{b\prime}} =  ({\rm Synth}_{{\rm PS1}^{b1}} + \Delta_{{\rm PS1}^{b1}}) - \\
    ({\rm Synth}_{S^{b\prime}} + \Delta_{S^{b\prime}}) + \\
    C_S^{b\prime}\times(({\rm Obs}_{{\rm PS1}^{b2}} + \Delta_{{\rm PS1}^{b2}}) - \\
    ({\rm Obs}_{{\rm PS1}^{b3}} + \Delta_{{\rm PS1}^{b3}}))~ \\
\end{aligned}
\label{eq:compare}
\end{equation}
{ where we define the observed stellar magnitude differences (left-hand-side of Equation~\ref{eq:compare}) as $\mathcal{R}_{{\rm Obs}^{b\prime}}$ and we define the differences from spectrophotometry of synthetic standards of Equation~\ref{eq:compare}) including propagating our fitted magnitude offsets for each filter (right-hand-side) as $\mathcal{R}_{{\rm Synth}^{b\prime}}$}. 

The vector of pre-computed synthetic magnitudes from CALSPEC and NGSL standards are ${\rm Synth}_{{\rm PS1}^{b1}}$ for PS1 bands and ${\rm Synth}_{S^{b\prime}}$ for each survey filter and the transformation from synthetic magnitudes is determined from color slope for a specific survey filter $C_S^{b\prime}$. The observed overlapping tertiary standards are used as follows. PS1 Public tertiary colors are computed as ${\rm Obs}_{{\rm PS1}^{b2}}-{\rm Obs}_{{\rm PS1}^{b3}}$. The synthetic transformation combined with the observed tertiary PS1 colors thus facilitates the comparison between observed overlapping tertiary star magnitudes between each survey filter being calibrated ${\rm Obs}_{S^{b\prime}}$ and the closest PS1 Public filter ${\rm Obs}_{{\rm PS1}^{b1}}$.}

{\red The offsets for PS1 Public filters} ($\Delta_{{\rm PS1}^{b1}}$, $\Delta_{{\rm PS1}^{b2}}$, and $\Delta_{{\rm PS1}^{b3}}$) and the specific survey filter ($\Delta_{S^{b\prime}}$) are floated with priors (described below) and are fit simultaneously. {\red Floating the PS1 Public offsets facilitates a simultaneous solution and covariance between all assessed filters.}  There are 105 filters analyzed here, 4 of which are the public PS1 cross-calibration filters in $griz$.  There are therefore 101 equations that are written in the form of Eq. \ref{eq:compare}, and 105 parameters. Degeneracies between parameters are broken with survey calibration priors explained further below.

In SuperCal, there was the option to fit linear transformation transformations ($C_S$) separately for the observed and synthetic sequences, thereby attempting to discern possible discrepancies in the mean wavelength of the effective filter.  Due to the complexity of the simultaneous fit, we fix the slope to that of the synthetic sequence, for survey-filters that were found to have consistent slopes between data and synthetic. Systematic uncertainties due to this decision are discussed in Section \ref{sec:discussion}. {\blue The only band with $>3\sigma$ difference in data/synthetic slopes was PS1 $g$ band when comparing to SDSS $g$ (+5$\sigma$), SNLS $g$ (+5.5$\sigma$), CfA1 $B$ (+3.5$\sigma$), CfA3S $B$ (+9$\sigma$), CfA4p1 $B$ (+5$\sigma$), KAIT $B$ (+4.5$\sigma$), SOUSA $B$ (+7$\sigma$). This was mediated by shifting the PS1 $g$ filter transmission by $+30$\AA\ to bring all into better agreement. {\red This shifting of the filter transmission is a correction on top of what was done in \cite{Tonry12}. \cite{Tonry12} performed a polynomial correction on the entire PS1 throughput to match the predicted spectro-photometry of the HST Calspec standards with their observed PS1 magnitudes.} }

We fit a solution that minimizes the $\chi^2$ differences between observed magnitudes from PS1 filters and each survey-filter following:
\begin{equation}
\chi^2 = \sum_{k=1}^{101} \frac{\langle\mathcal{R}_{{\rm Obs}^{b\prime}}-\mathcal{R}_{{\rm Synth}^{b\prime}}\rangle^2}{\sigma_{S^{bk}}^2/N + f_{S^{bk}}^2} \times W_S + \frac{1}{2}(\frac{\Delta_{S^{bk}}}{\sigma_{p^k}})^2
\label{eq:chisq}
\end{equation}
where the summation is over $k$-bands {\red which is 101 when not including the 4 public PS1 bands used to perform the comparisons themselves}, and $N_t$ overlapping tertiaries. The uncertainties are determined from the {\red photometric scatter $\sigma_{S^{bk}}$} and the designated error floor $f_{S^{bk}}$. Priors ($\sigma_{p^k}$) centered at zero offset are included for each survey-filter. We use the \texttt{emcee} MCMC sampling library described by \cite{emcee} to minimize $\chi^2$ (Equation \ref{eq:chisq}) and to obtain an uncertainty covariance matrix.
    
{\blue There are three additional components of Eq.~\ref{eq:chisq} required to compute a robust solution and uncertainties: 

\textbf{1)} Survey-filter zeropoint priors ($\sigma_{p^k}$) representing the prior knowledge/confidence of the original calibration of each system (after updating their calibrations to the latest CALSPEC as shown in Appendix Table \ref{tab:recal}). Priors are centered at zero. For the rolling SN surveys with modern and reliable calibrations (PS1/Foundation, DES, SDSS, and SNLS) we utilize priors of $0.00\pm0.01$~mag. {\blue We do not apply zeropoint priors to any other samples as to ensure that the calibration solution is primarily determined by the more modern all sky surveys.} Unlike SuperCal, which fixed the calibration of the public PS1 catalog ($\sigma_{p^k}=0$) used to perform the cross calibration, here we place conservative priors of width of 0.02  mag to facilitate covariance between all of the surveys in the fitting process.

\textbf{2)} Relative survey weights ($W_S$) to account for photometric systems that appear multiple times but are the same telescope having been recalibrated over time (e.g. KAIT 1,2,3,4 receive a $W_S=1/4$ in Eq.~\ref{eq:chisq})

\textbf{3)} Photometric uncertainty floors ($f_{S^{bk}}$) to account for the observed scatter of the brightest tertiary standards that is not explained by photometric uncertainties alone. PS1/Foundation, DES, SDSS, and SNLS are given photometric uncertainty floors ($f_{S^{bk}}$) of 0.005 and the remaining low-$z$ surveys are given 0.01 (except for CNIa0.02 for which we did not have tertiary stars we apply a 0.02 mag floor).
 }

\subsection{Data Preparation}
For each survey we match astrometric positions of the tertiary catalogs with those in PS1 to within $<$ 1 arcsec. We avoid potential errors from blending by only choosing isolated stars with no other star ($m < 22$ mag) within 15 arcsec.

For each of the stars in the spectral libraries we integrate the spectrum with the throughput of each of the passbands of the individual surveys to determine synthetic magnitudes. To facilitate comparison between the observed sequence of stars and the synthetic magnitudes, we correct the observed stellar magnitudes for Milky Way extinction using the known positions of the stars. The extinction values and attenuation curves are queried from IRSA\footnote{\href{https://irsa.ipac.caltech.edu/}{https://irsa.ipac.caltech.edu/}} for the maps from \cite{Schlafly11} and are interpolated at the mean wavelength of each filter.

For comparison with robust PS1 magnitudes, we adopt a brightness cut on the PS1 magnitudes eliminating stars with magnitudes brighter than [14.8,14.9.15.1,14.6] in [g,r,i,z] because of the concerns of non-linearity noted in \cite{Schlafly12}. For the PS1 star catalog we also place a cut requiring that the PS1 $g$ is brighter than 19th magnitude in order to avoid Malmquist bias in the tertiary star selection.  Lastly, we choose a specific color range ($0.25 < g - i < 1.0$) of the PS1 catalog stars used in the analysis {\blue to maximize statistics, minimize dispersion, and insure linearity of the transformation in the synthetic library} in the region of comparison for color transformations (the exceptions for this are the LOSS-Nickel and CNIa0.02 datasets for which we use narrow color ranges, as discussed above in Sections~\ref{sec:loss}~\&~\ref{sec:cnia}).

\subsection{Calibration Solution Results}

\begin{table}
\centering
\resizebox{.5\textwidth}{!}{
\centering
\begin{tabular}{llc}
\toprule
    SURVEY & FILTER &           OFFSET \\
\hline
       CFA1 &    $  B $&  NA \\
       CFA1 &    $  V $&  NA \\
       CFA1 &    $  R $&  NA \\
       CFA1 &    $  I $&  NA \\
       CFA2 &    $  B $&  NA \\
       CFA2 &    $  V $&  NA \\
       CFA2 &    $  R $&  NA \\
       CFA2 &    $  I $&  NA \\
      CFA3S &    $  B $&  -0.038$\pm$0.012 \\
      CFA3S &    $  V $&  -0.009$\pm$0.010 \\
      CFA3S &    $  r $&  -0.014$\pm$0.011 \\
      CFA3S &    $  i $&   -0.005$\pm$0.011 \\
      CFA3K &    $  B $&  -0.025$\pm$0.012 \\
      CFA3K &    $  V $&  -0.006$\pm$0.010 \\
      CFA3K &    $  r $&    0.001$\pm$0.012 \\
      CFA3K &    $  i $&  0.005$\pm$0.010 \\
      CFA4p1 &   $  B $&   -0.006$\pm$0.012 \\
      CFA4p1 &   $  V $&    -0.001$\pm$0.010 \\
      CFA4p1 &   $  r $&  0.003$\pm$0.011 \\
      CFA4p1 &   $  i $&  0.005$\pm$0.012 \\
      CFA4p2 &   $  B $&   0.028$\pm$0.011 \\
      CFA4p2 &   $  V $&   0.005$\pm$0.011 \\
      CFA4p2 &   $  r $&   0.010$\pm$0.010 \\
      CFA4p2 &   $  i $&   0.010$\pm$0.011 \\
     CSPDR3 &    $  g $&   0.000$\pm$0.011 \\
     CSPDR3 &    $  r $&    -0.011$\pm$0.010 \\
     CSPDR3 &    $  i $&  -0.024$\pm$0.010 \\
     CSPDR3 &    $  B $&  -0.015$\pm$0.011 \\
     CSPDR3 &    $  V $&  -0.02$\pm$0.011 \\
     CSPDR3 &    $  V$1 &  -0.019$\pm$0.011 \\
     CSPDR3 &    $  V$2 &  -0.022$\pm$0.011 \\
     CSPDR3 &    $  V$3 &  -0.011$\pm$0.011 \\
    KAIT1 Ganesh &   $   B$ &   0.001$\pm$0.012 \\
    KAIT1 Ganesh &   $   V$ &   0.005$\pm$0.011 \\
    KAIT1 Ganesh &   $   R$ &  -0.001$\pm$0.010 \\
    KAIT1 Ganesh &   $   I$ &   0.005$\pm$0.011 \\
    KAIT2 Ganesh &   $   B$ &   -0.002$\pm$0.012 \\
    KAIT2 Ganesh &   $   V$ &   0.006$\pm$0.010 \\
    KAIT2 Ganesh &   $   R$ &   -0.001$\pm$0.011 \\
    KAIT2 Ganesh &   $   I$ &    -0.000$\pm$0.010 \\
    KAIT3 Ganesh &   $   B$ &  -0.002$\pm$0.013 \\
    KAIT3 Ganesh &   $   V$ &   0.005$\pm$0.011 \\
    KAIT3 Ganesh &   $   R$ &  -0.000$\pm$0.010 \\
    KAIT3 Ganesh &   $   I$ &  -0.000$\pm$0.011 \\
    KAIT4 Ganesh &   $   B$ &  -0.000$\pm$0.012 \\
    KAIT4 Ganesh &   $   V$ &     0.006$\pm$0.010 \\
    KAIT4 Ganesh &   $   R$ &    -0.001$\pm$0.010 \\
    KAIT4 Ganesh &   $   I$ &  -0.000$\pm$0.011 \\
  NICKEL1 Ganesh &   $   B$ &   0.004$\pm$0.010 \\
  NICKEL1 Ganesh &   $   V$ &   0.005$\pm$0.010 \\
  NICKEL1 Ganesh &   $   R$ &  -0.001$\pm$0.010 \\
  NICKEL1 Ganesh &   $   I$ &   0.0017$\pm$0.011 \\
  NICKEL2 Ganesh &   $   B$ &  -0.046$\pm$0.013 \\
  NICKEL2 Ganesh &   $   V$ &   0.005$\pm$0.010 \\
  NICKEL2 Ganesh &   $   R$ &  -0.003$\pm$0.010 \\
  NICKEL2 Ganesh &   $   I$ &   0.029$\pm$0.010 \\

\hline

\end{tabular}

\begin{tabular}{llc}
\toprule
    SURVEY & FILTER &           OFFSET \\
\hline
     KAIT3 Stahl &      $B$ &   0.019$\pm$0.015 \\
     KAIT3 Stahl &      $V$ &   0.047$\pm$0.026 \\
     KAIT3 Stahl &      $R$ &   0.041$\pm$0.022 \\
     KAIT3 Stahl &      $I$ &   0.028$\pm$0.015 \\
     KAIT4 Stahl &      $B$ &   0.016$\pm$0.011 \\
     KAIT4 Stahl &      $V$ &   0.026$\pm$0.010 \\
     KAIT4 Stahl &      $R$ &   0.017$\pm$0.011 \\
     KAIT4 Stahl &      $I$ &   0.006$\pm$0.011 \\
   NICKEL1 Stahl &      $B$ &   0.014$\pm$0.012 \\
   NICKEL1 Stahl &      $V$ &   0.025$\pm$0.012 \\
   NICKEL1 Stahl &      $R$ &   0.018$\pm$0.010 \\
   NICKEL1 Stahl &      $I$ &   0.012$\pm$0.012 \\
   NICKEL2 Stahl &      $B$ &  -0.053$\pm$0.012 \\
   NICKEL2 Stahl &      $V$ &   0.027$\pm$0.011 \\
   NICKEL2 Stahl &      $R$ &   0.013$\pm$0.011 \\
   NICKEL2 Stahl &      $I$ &    0.040$\pm$0.011 \\
           SOUSA &      $B$ &  -0.013$\pm$0.012 \\
           SOUSA &      $V$ &   0.027$\pm$0.011 \\
        CNIa0.02 LCO &      $B$ &  -0.026$\pm$0.022 \\
        CNIa0.02 LCO&      $V$ &  -0.034$\pm$0.021 \\
        CNIa0.02 LCO&      $r$ &   -0.005$\pm$0.020 \\
        CNIa0.02 LCO&      $i$ &   0.011$\pm$0.021 \\
        CNIa0.02 RP&    $B$ &  -0.068$\pm$0.020 \\
        CNIa0.02 RP&      $V$ &  -0.022$\pm$0.022 \\
        CNIa0.02 RP&      $r$ &  -0.000$\pm$0.021 \\
        CNIa0.02 RP&      $i$ &   0.017$\pm$0.021 \\
         DES 3YR &      $g$ &   0.003$\pm$0.007 \\
         DES 3YR &      $r$ &   -0.015$\pm$0.005 \\
         DES 3YR &      $i$ &  -0.005$\pm$0.006 \\
         DES 3YR &      $z$ &  -0.002$\pm$0.006 \\
         DES 5YR &      $g$ &   0.002$\pm$0.006 \\
         DES 5YR &      $r$ &  -0.009$\pm$0.006 \\
         DES 5YR &      $i$ &    -0.007$\pm$0.006 \\
         DES 5YR &      $z$ &   0.006$\pm$0.006 \\
            SDSS &      $g$ &  -0.008$\pm$0.006 \\
            SDSS &      $r$ &   0.004$\pm$0.005 \\
            SDSS &      $i$ &   0.003$\pm$0.006 \\
            SDSS &      $z$ &    -0.005$\pm$0.006 \\
            SNLS &      $g$ &   0.001$\pm$0.005 \\
            SNLS &      $r$ &  -0.002$\pm$0.005 \\
            SNLS &      $i$ &   0.002$\pm$0.005 \\
            SNLS &      $z$ &   0.000$\pm$0.006 \\
      PS1 Public &      $g$ &  -0.010$\pm$0.003 \\
      PS1 Public &      $r$ &   0.005$\pm$0.003 \\
      PS1 Public &      $i$ &    0.002$\pm$0.004 \\
      PS1 Public &      $z$ &    0.013$\pm$0.004 \\
         PS1 SNe &      $g$ &   -0.023$\pm$0.006 \\
         PS1 SNe &      $r$ &   -0.025$\pm$0.006 \\
         PS1 SNe &      $i$ &  -0.018$\pm$0.006 \\
         PS1 SNe &      $z$ &  -0.013$\pm$0.006 \\
      Foundation &      $g$ &  -0.007$\pm$0.006 \\
      Foundation &      $r$ &   0.008$\pm$0.006 \\
      Foundation &      $i$ &    0.004$\pm$0.007 \\
      Foundation &      $z$ &   0.013$\pm$0.007 \\

        &       &    \\
        &       &    \\
\hline

\end{tabular}

}
\caption{Fragilistic best-fit calibration solution. The individual survey-filter offsets and uncertainties are given in mags. }
\label{tab:offsets}
\end{table}

   \begin{figure}
        \centering 
	    \includegraphics[width=.48\textwidth]{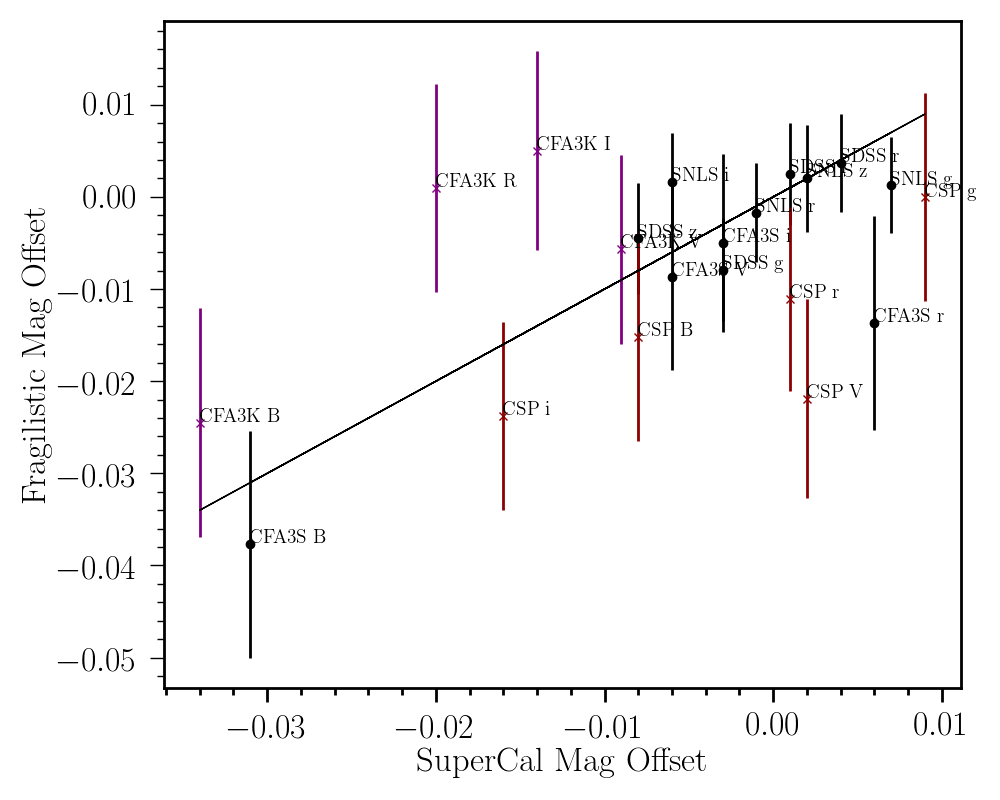} 
        \caption{{\blue The Fragilisitc offsets found here for various filters and systems compared to those} found in SuperCal.  The most significant differences are due to the new definition of CSP bands (shown in red) {\blue and CfA3K bands (shown in purple)}. {\red Black line is y=x.}}
        \label{fig:SuperCal} 
    \end{figure}

The best-fit offsets for each of the 105 filters are given in Table~\ref{tab:offsets}.  Overall, for the subset of filters that were also calibrated in SuperCal, we find relatively good agreement (Fig.~\ref{fig:SuperCal}) and the observed differences between Fragilistic and SuperCal are either traced to changes in the filter throughputs, {\blue the reliance of surveys on specific CALSPECs that have been redefined}, or the novel inclusion of individual filter uncertainties in the simultaneous fit. The largest difference we find is for CSP $V$ band, which due to numerous changes by the CSP collaboration had different filter transmissions defined for use in this paper and SuperCal.  {\red The derived offsets upon journal acceptance will made available in machine readable format\footnote{\href{https://github.com/PantheonPlusSH0ES/DataRelease/}{https://github.com/PantheonPlusSH0ES/DataRelease/}}.}

    \begin{figure*}
        \centering 
	    \includegraphics[width=1.05\textwidth]{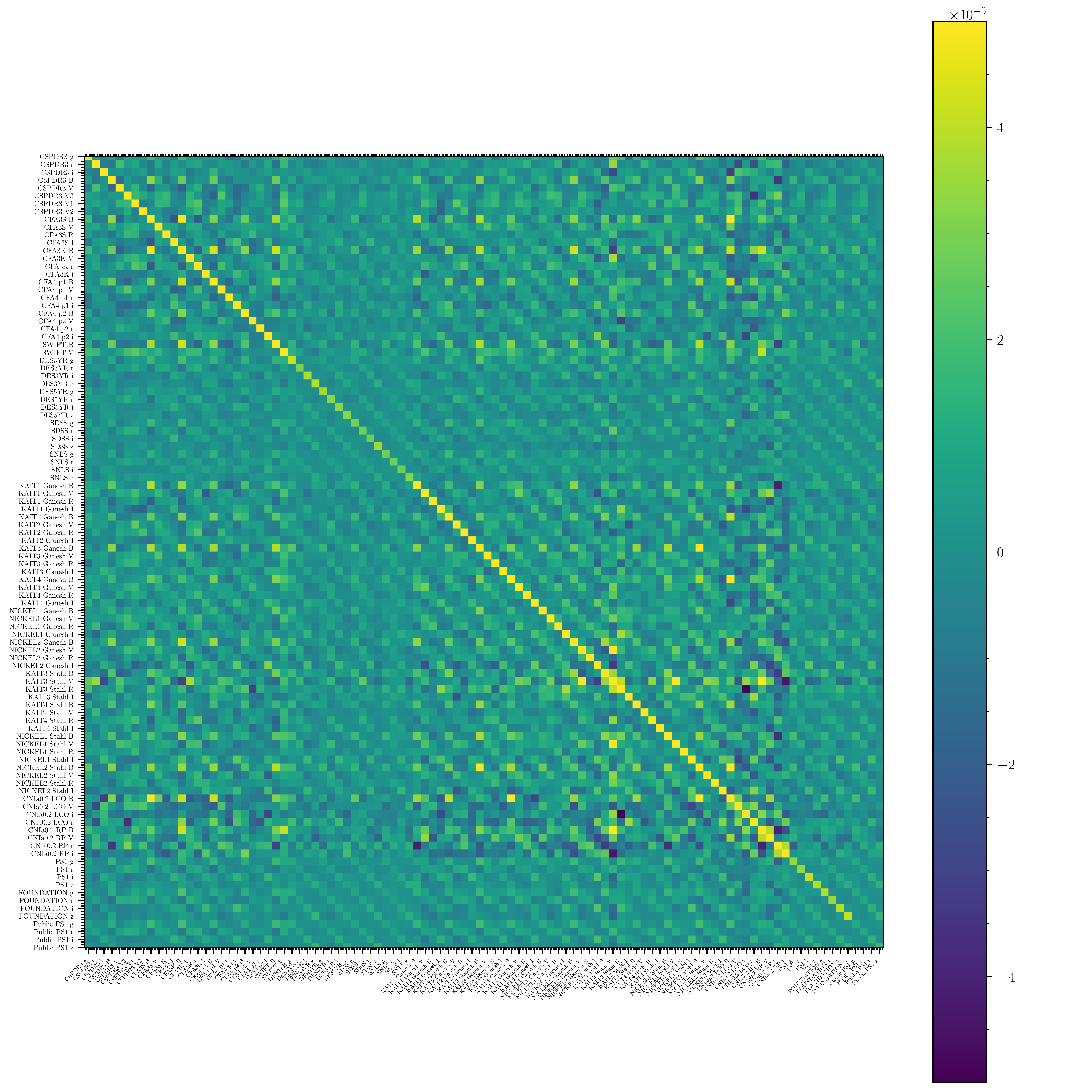} 
        \caption{The covariance matrix between fitted Fragilistic zeropoint offsets of all the filters used in this analysis.  This matrix can be downloaded here: \href{https://github.com/PantheonPlusSH0ES/DataRelease/}{https://github.com/PantheonPlusSH0ES/DataRelease/}}
        \label{fig:fragilisticcov} 
    \end{figure*}

The covariances between the filter zeropoints are shown in Figure~\ref{fig:fragilisticcov}. The covariance matrix will be downloadable in machine readable format$^7$. The diagonal terms of the covariance, the statistical uncertainties on the calibration offsets, are typically bounded by the error floors given for each survey. Off-diagonal covariance terms between different surveys are largest for the newer surveys with limited calibration stars (i.e. between CNIa0.02 and LOSS or SOUSA). Low-$z$ survey $B$ band constraints are also often amongst those with the largest covariances because they have minimal overlap with PS1 bands used for cross-calibration.

   \begin{figure}
        \centering 
	    \includegraphics[width=.48\textwidth]{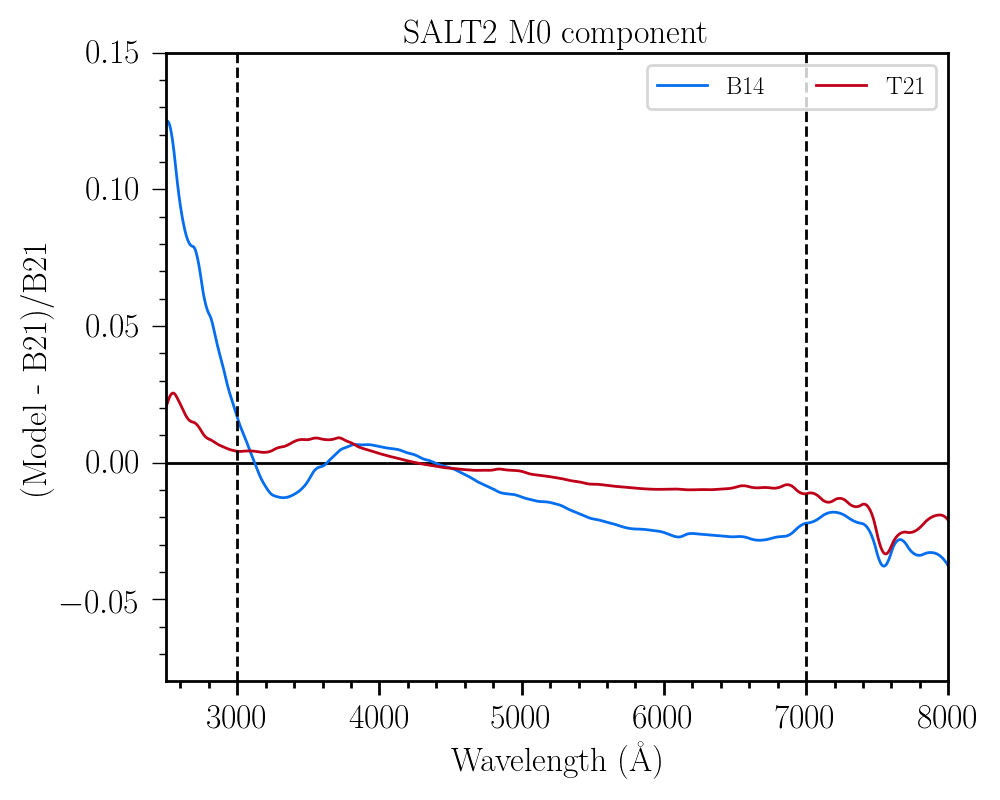} 
        \includegraphics[width=.48\textwidth]{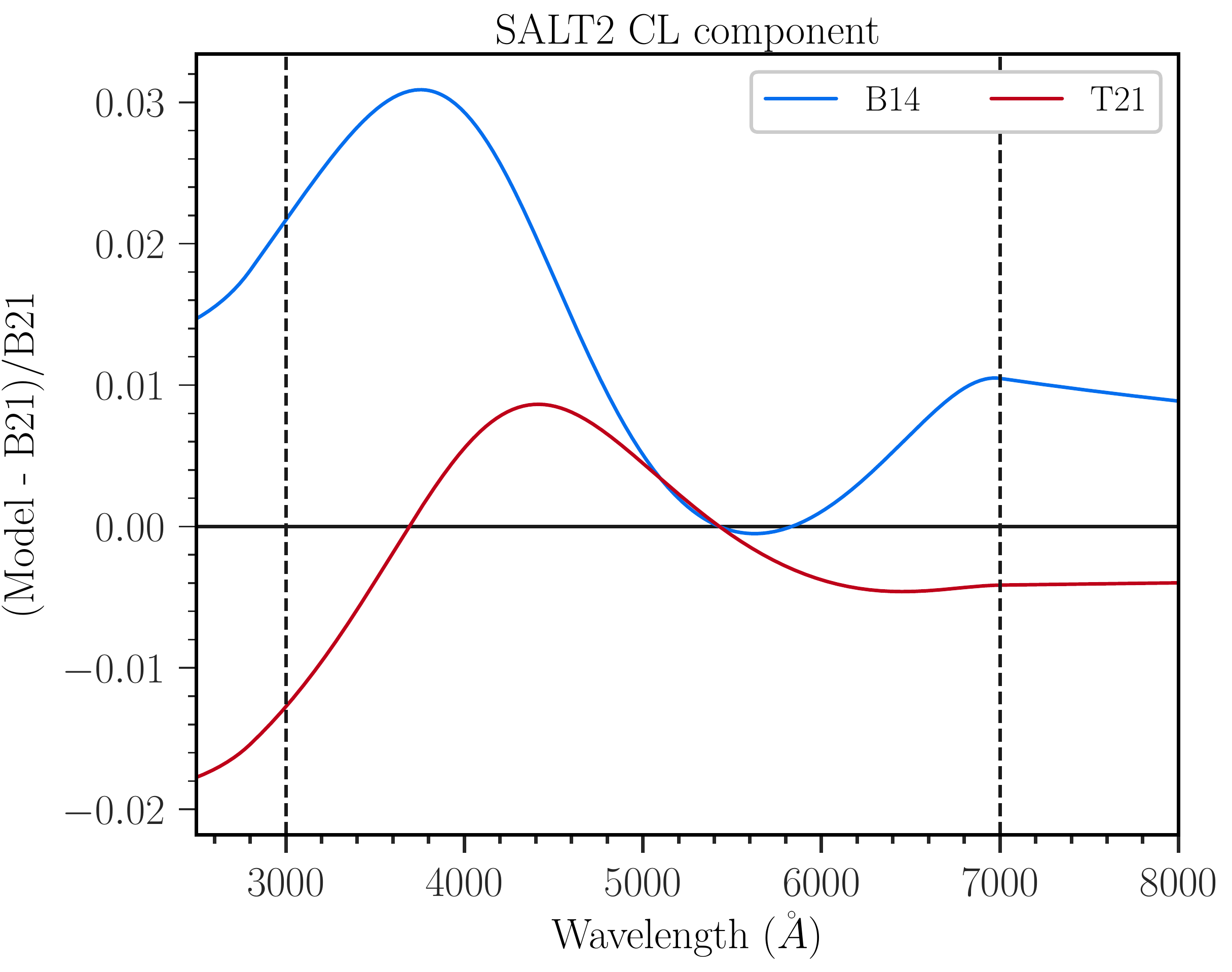} 
\caption{[Top] {\blue The difference in SALT2 M0 component for B14, T21, and B21 as a function of wavelength. Vertical dashed lines indicate the allowed SALT2 wavelength range used for light-curve fitting.} [Bottom] The difference in the SALT2 color law component CL.}
        \label{fig:saltsurfb14} 
    \end{figure}

\section{Re-Training the SALT2 model}

{\blue Given a new calibration for the samples of light-curves} used in SALT2 training, as demonstrated in \cite{Taylor21}, one should retrain the SALT2 model (to avoid model calibration systematics biasing the resulting SALT2 light-curve fit results). \cite{Taylor21} establishes the up-to-date infrastructure to do so. We note that most previous analyses \citep{Scolnic18,Brout18b,Jones18} assumed a large additional systematic uncertainty from the fact that they did not retrain the SALT2 surface. Here we describe both the retraining as well as the systematic uncertainty that arises from the cross-calibration solution described by the Fragilistic covariance matrix.

\subsection{Training of the Nominal SALT2 model with the Fragilistic Calibration Solution.}

With the new calibration zeropoints determined here and the updated filter functions, we retrain the SALT2 model following \cite{Taylor21} which uses the algorithms established in \cite{Guy2010}.  We denote our newly trained SALT2 surface `B21'. Previously used surfaces include the model trained for the use in the Joint Light Curve Analysis but also used in the original Pantheon analysis, denoted `B14', and the model trained in \cite{Taylor21} on the original SuperCal calibration solution, denoted `T21'.  {\red While we have not re-calibrated any rest-frame $u$ band filters due to lack of overlap with PS1 wavelengths, in the training of SALT2 $u$ band is used for SDSS and CSP with recalibration zeropoint offsets set to zero. We note the impact of this choice in Section \ref{sec:discussion} and Appendix~\ref{sec:nou}.}

We present the fractional differences between B21, T21, and B14 in the average spectral energy density for a fiducial (x1=0, c=0) SN Ia (M0) and color law (CL) in Fig.~\ref{fig:saltsurfb14}. The rest-frame wavelength bounds (3000A, 7000A), within which the model is used for Pantheon+ light-curve fits, are shown by vertical dashed lines. The largest differences occur in the M0 surface component, particularly at the UV end, although not over the region that is used in the light-curve fitting. {\blue Over the $g-z$ color range (3000\AA~-~7000\AA) we find a slope of $\sim2.5$\% between B21 and B14 and a slope of $\sim1.5$\% between B21 and T21.}

\subsection{Distances and Cosmology}

We evaluate the impact of our Fragilistic calibration and newly retrained SALT2 model by comparing fits of SN light-curve photometry for the three aforementioned SALT2 models (B21, T21, and B14). 
To determine distances we use the traditional \cite{Tripp98} estimator:

\begin{equation}
\label{Eq:tripp}
  \mu = m_B - M + \alpha x_1 - \beta c 
\end{equation}
where $m_b$ is the converted peak-magnitude of SNe Ia from the fitted flux normalization $x_0$, $M$ is the fiducial absolute magnitude of a Type Ia SN, $x_1$ is the stretch parameter, $c$ is the color parameter, and $\alpha$ and $\beta$ are correlation coefficients {\blue that minimize the scatter in the standardized luminosities. We fit a flat $w$CDM cosmological model to the observed distances. }

To better understand the sensitivity of the re-training to the new calibration, we performed two sets of tests. First, we first fixed the M0 surface to that of B14 and only replaced the color law (CL) with the fitted CL for B21. We then performed the opposite and change only the M0 surface to B21 and keep the CL fixed to that of B14.  We find that the sensitivity to cosmology, i.e. redshift dependence of inferred distances, is dominated by the M0 surface. 

\begin{figure}
        \centering 
	    \includegraphics[width=.48\textwidth]{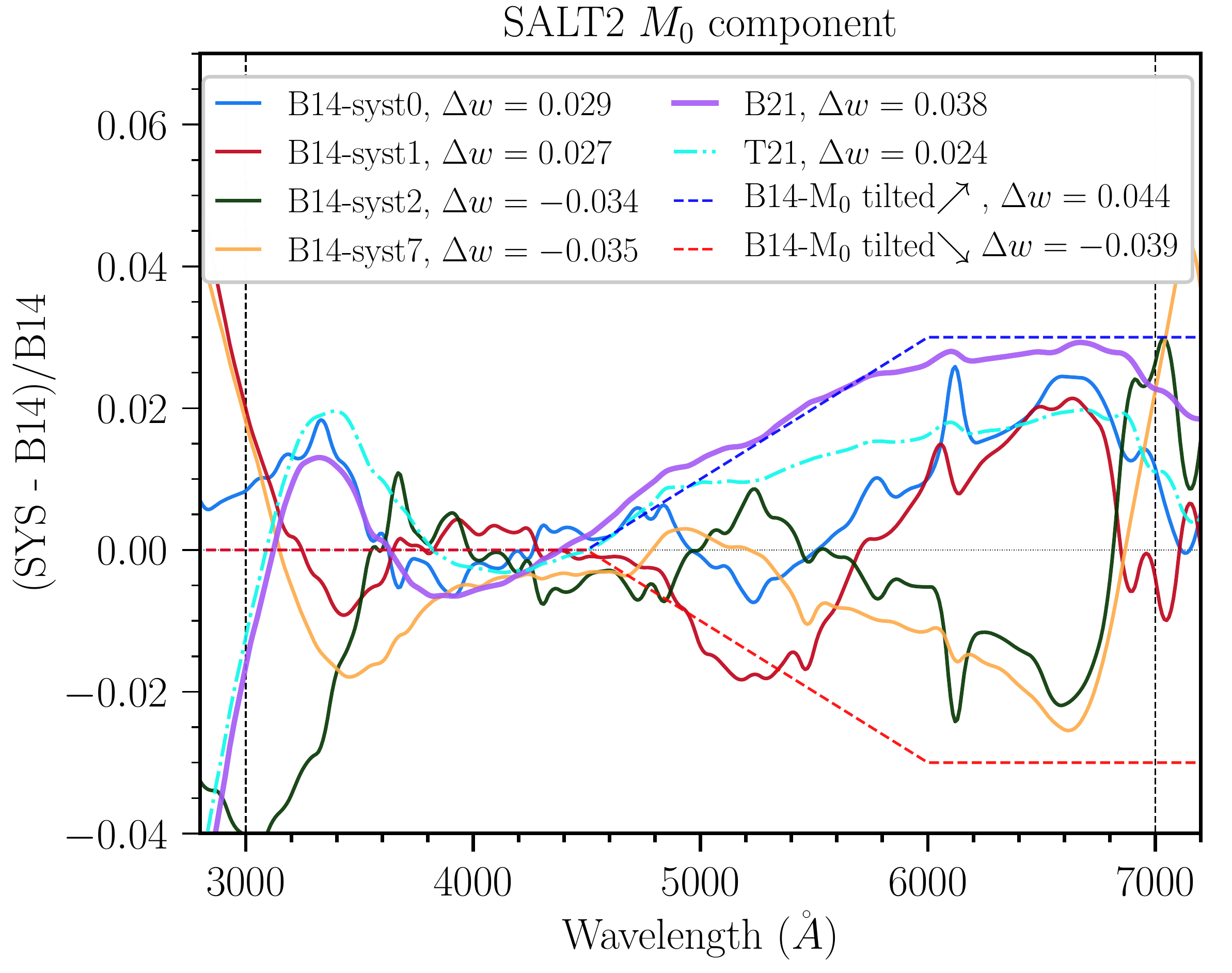} 
\caption{{\blue Same as Figure~\ref{fig:saltsurfb14} but for several different retrained SALT2 models. Constraints on dark energy equation of state $w$ are shown with a prior of 0.3$\pm$0.01 on $\Omega_M$}}
        \label{fig:m0cosmology} 
    \end{figure}

The slope in the M0 component seen in Fig.~\ref{fig:saltsurfb14} directly results in redshift-dependent effects on distance (due to different observer frame filters being used), and thus a systematic sensitivity to cosmological parameters. In Figure~\ref{fig:m0cosmology}, the percentage differences in M0 are shown for several SALT2 models including several systematic surfaces B14-syst* provided by \cite{Betoule2014}. We also design by hand M0 surfaces that exhibit fixed slopes (both positive and negative; `tilted up' and `tilted down') relative to B14 over the wavelength range 4500 to 6000 and propagate all of these models to cosmological parameter inference. We find that recovered $w$ values clearly correlate both with the size of the slope and the direction of the slope in wavelength and are confirmed by our `tilted' surfaces. The largest cause of the change to the M0 surface can be traced to the recent update of the fundamental calibration of CALSPEC standards in \cite{Bohlin21}. 

For the second set of tests we perturb each filter in the SALT2 training individually and compute the resulting sensitivity to the fitted cosmological parameter $w$. The top five most sensitive components for measurements of $w$ are shown in Table~\ref{tab:sensitivity}. All of the most sensitive perturbations are related to filter zeropoints even though we also performed perturbations to the effective mean wavelengths of each filter in the training.

\begin{table}
\centering
\begin{tabular}{lrcr}
\toprule
 Survey & Filter & $\delta w/ 0.01 {\rm mag}$ \\
\hline
            SDSS &  g  &  -0.0095\\
 CfA3 Keplercam  &  V  &  -0.0087\\
         SNLS    &  r  &  -0.0060\\
         SNLS    &  i  &   0.0042\\
   CfA3 Standard &  B  &  -0.0036\\
    CfA3 Standard&  R  &  -0.0031\\
        CSP      &  u  &  -0.0027\\
         SDSS    &  u  &  -0.0025\\
   CfA3 Standard &  U  &  -0.0024\\
  CfA3 Standard  &  V  &  -0.0024\\
\hline

\end{tabular}
\caption{{\blue Filter zeropoint sensitivity to cosmological constraints ($\delta w/0.01 {\rm mag}$) for 0.01mag perturbations in SALT2 training. Only the top 10 most sensitive perturbations are shown. Sensitivity to mean wavelengths of each filter was examined but none are within top 10 most sensitive.}
}
\label{tab:sensitivity}
\end{table}

\begin{table*}
\centering
\begin{tabular}{llrr}
\toprule
    Systematic & Description & Size $\sigma_k$ & $\sigma_w$\\
\hline
      CALSPEC Modeling & 0.005mag/$7000\AA$&    3 & 0.005\\
      Survey Cal. & 9 correlated realizations &   1/3 each & 0.010\\
       & of SALT2 and zeropoints &   &\\
      Extra SALT2  & B14 Surface & 1/3 & 0.007\\
      Extra CSP Cal. & \cite{Stritzinger2010} Stars &  1/2 & 0.001\\
\hline
      Total & &  & 0.013\\
\end{tabular}
\caption{{\blue The systematic uncertainties associated with calibration for the Pantheon+ analysis (Brout et al. in prep) and their scaled contribution in the building of the covariance matrix ($\sigma_k$ of Eq.~\ref{eq:cov}). The uncertainty on CALSPEC modeling has been tripled ($\sigma_k=3$) in comparison to the original Pantheon \citep{Scolnic18}. Extra SALT2 and CSP Cal. refer to the fact that SALT2 and CSP are both already included in Survey Cal., but in order to be conservative we include extra sources of systematics. Systematic uncertainties on dark energy ($\sigma_w$) are computed when combining with Planck \cite{planck}.} }
\label{tab:sysdescription}
\end{table*}

We note that bias correction simulations using the SALT2 model are often used to correct observables by expected biases. The impact of changing bias correction simulations with the new SALT2 model is $<0.1\%$ in $w$. The reason for this small impact is that in the determination of the bias correction sample, the same SALT2 model is used in the simulation and the fitting. {\blue This is unlike the impact on the data itself, which does not benefit from this same cancellation and therefore it is only the fitting of the data that dominates the impact of changing SALT2. We therefore do not include bias corrections in estimating the impact of systematic uncertainties here.}  

\subsection{Systematic Uncertainty}
\label{sec:sys}
In order to quantify the impact of calibration uncertainties on SN distances, we develop a novel technique in which we both retrain SALT2 model and apply zeropoint offsets in the light-curve fitting simultaneously. We utilize the cross-calibration covariance matrix obtained in the Fragilistic MCMC fitting process (Fig. \ref{fig:fragilisticcov}). To do so we perform a Cholesky decomposition of the covariance matrix to create 9 mock realizations of calibrations with correlated survey zeropoints as well as with uncorrelated effective mean wavelength shifts of each filter used in training (see Appendix Table~\ref{tab:wavunc}). We then train 9 SALT2 surfaces on the realizations and the resulting 9 SALT2 M0 components and color laws are shown in Appendix Figure~\ref{fig:saltsurfsys}. The fractional differences between the the 9 systematic variant M0 components and the nominal component corresponding to the best fit Fragilistic offsets (horizontal line) resemble in scale the differences seen between the original SALT2 calibration (B14) and Fragilistic (Appendix Fig. \ref{fig:saltsurfsys}).  

Converting the covariance of Fig. \ref{fig:fragilisticcov} which is in filter magnitude space into a distance-based covariance is non-trival from first principles. Instead, to compute a distance covariance that can be used to constrain cosmological models, we apply the same set of mock zeropoint offsets in the light-curve fitting and propagate the differences in both the 9 SALT2 surfaces and zeropoints to differences in cosmological distances. This is shown in Appendix Fig. \ref{fig:salt2retrainmudif}. Finally, following \cite{Conley2010}, we propagate the resulting set of 9 Hubble diagrams to a distance $\times$ distance covariance matrix.

\begin{equation}
C_{{\zindex}_i{\zindex}_j} = \sum_{k} \frac{\partial \Delta\mu_{{\zindex}_i}}{\partial k} ~ \frac{\partial \Delta\mu_{{\zindex}_j}}{\partial k} ~ \sigma_k^2,
\label{eq:cov}
\end{equation}
where the summation is over the systematics ($k$), $\Delta\mu_{{\zindex}_i}$ {\blue are the residuals in distance for the SNe fitted between each SALT2 model}, and $\sigma_k$ is 1/3 such that when the 9 systematic vectors are added in quadrature, they sum to $\sim1$. The resulting matrix is utilized in cosmology fitting as discussed in Section~\ref{sec:syscosmo}.

{\blue In addition, we build covariance matrices for several other calibration-related systematics. This is done following Eq.~\ref{eq:cov} where we sum over each systematic perturbation to the analysis described in Table~\ref{tab:sysdescription}. As there have been numerous significant updates to CALSPEC over the years as described in Section~\ref{sec:CALSPEC}, we adopt a 3$\times$ larger systematic uncertainty than is described in \cite{Bohlin21} and than was adopted in the original Pantheon cosmological analysis. Additionally, we adopt a systematic of 1/3 the difference in distances derived between B14 and B21 SALT2 surfaces, in order to conservatively account for possible systematic from the SALT2 model training process. Lastly we include an additional conservative systematic uncertainty due to the re-calibration of the CSP tertiary standard stars.}

\section{Agreement of Survey Hubble Residuals and Impact on Cosmology}
\label{sec:cosmology}

\subsection{Agreement of Survey Hubble Residuals}

\begin{figure*}
        \centering 
	    \includegraphics[width=.88\textwidth]{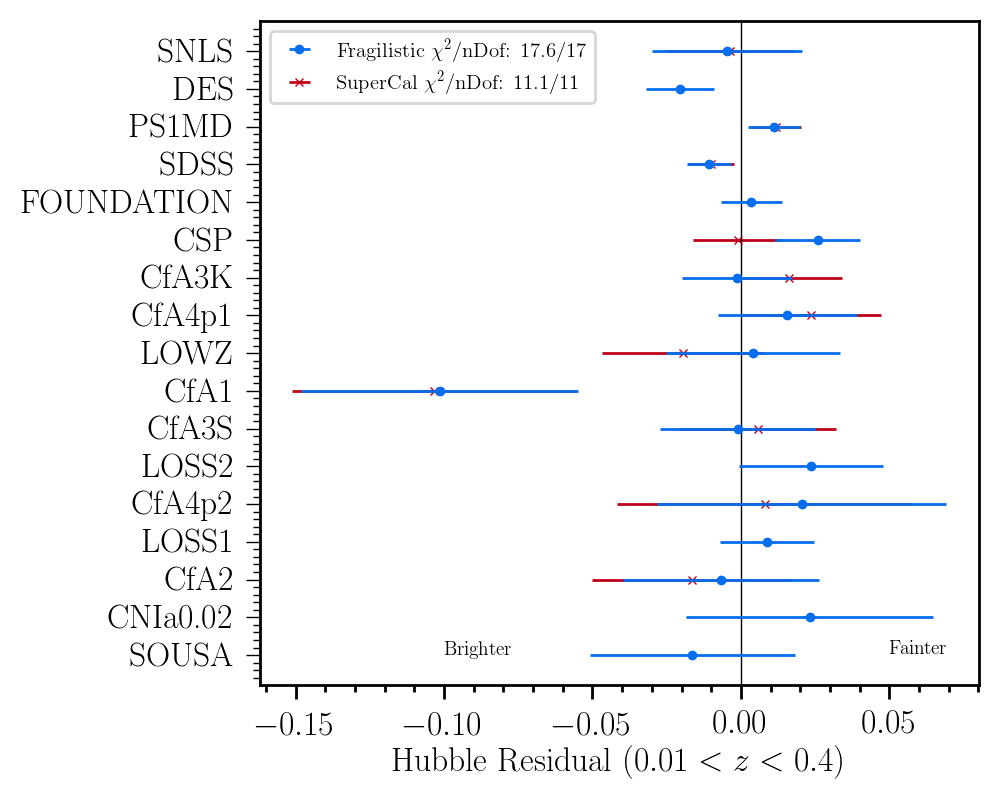} 
        \caption{ Weighted average Hubble residual offsets to the best-fit
$\Lambda$CDM model for SN observed by each survey. All SNe have been corrected for expected biases in observable parameters (described in Brout et al in prep). The
points in red represent the residuals using the SuperCal calibration \citep{Scolnic16}. The blue points represent the
offsets after applying Fragilistic calibration.  When no SuperCal correction is given, this is because this survey was not included in the SuperCal calibration. The ordering of surveys on the y-axis is in order of increasing sample mean redshift (from bottom to top).}
        \label{fig:survmeans} 
    \end{figure*}

To assess the level of improvement with the Fragilistic solution, we compare the survey offsets {\blue in the Hubble Diagram}. As shown in Fig.~\ref{fig:survmeans}, we calculate the weighted average of the Hubble residuals for each survey over a redshift range of $0.01<z<0.4$ relative to the best-fit cosmology ($\chi^2/N_{dof}=17.6/17$). We cut at a maximum redshift of 0.4 to eliminate sensitivity to cosmological signals. We also note that we do not expect fully independent distribution of survey offsets because there exist SNe that have beeen observed by multiple surveys. {\blue We also report the offsets for the subset of surveys calibrated in SuperCal ($\chi^2/N_{dof}=11.1/11$). We find good agreement with the original Pantheon sample and calibration with the exception of CSP which had its tertiary star catalog updated in \cite{krisciunas20} and for which we add additional systematic uncertainty (Last row of Table~\ref{tab:sysdescription}). Small differences in Hubble residuals for surveys that have not been recalibrated (i.e. CfA1) are due to changes in the SALT2 training. Surveys in Fig.~\ref{fig:survmeans} are ordered from top to bottom by mean survey redshift and no trends as a function of redshift are seen. We note that while the CfA1 survey offset shown in Fig.~\ref{fig:survmeans} is for the Hubble flow SN photometry from \cite{riess99}, the H$_0$ Cepheid calibrators from CfA1 (SN 1994ae and SN 1995al) were recalibrated in \cite{riess05,riess09}.}

\subsection{Impact on Distances and Cosmology}

\begin{figure}
        \centering 
	    \includegraphics[width=.48\textwidth]{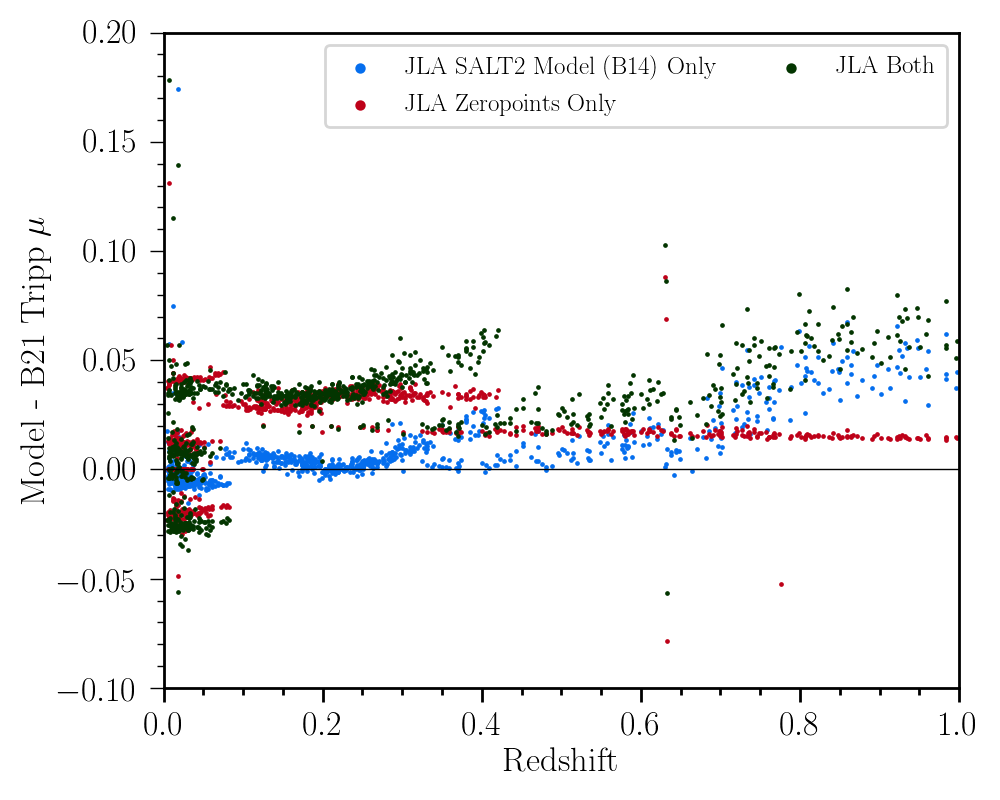} 
        \caption{The net change in the Tripp distance modulus values as given by Eq.~\ref{Eq:tripp} due to re-fitting  the light-curves with 1) new B21 calibration but while the keeping JLA SALT2 model (blue) 2) re-training the SALT2 B21 model but keeping JLA zeropoints for the fitting (red) and 3) neither re-training the SALT2 model nor updating calibration zeropoints (black). }
        \label{fig:salt2zdep} 
    \end{figure}

In Fig.~\ref{fig:salt2zdep}, we show the impact on recovered distances for the subset of $\sim700$ SNe that were calibrated for the JLA analysis \citep{Betoule2014}. We find a difference in $w$ of +0.035 using the B21 SALT2 surface presented in this work, versus the B14 surface used in the JLA/Pantheon analysis (when applied to the same data).  We can trace nearly the entirety of this shift to the change in the M0 component, rather than any other change in the components of the SALT2 model. We also find a difference in $w$ of +0.025 from the update to the Fragilistic calibration solution and its affect on the survey zeropoints in light-curve fitting.
Distance modulus residuals are shown in Fig.~\ref{fig:salt2zdep} relative to JLA for three cases: 1) calibration zeropoints (ZPT) changed to Fragilistic (no SALT2 retraining), 2) SALT2 retraining to B21 model (no calibration offsets changed), and 3) both Fragilistic zeropoints and SALT2 B21 retraining done simultaneously. Distances are shown as a function of redshift, $z$, in order to understand the impact on cosmological model constraints. For the SALT2 retraining only, we find a slope relative to B14 over a $\Delta z=1$ is $\sim0.04$ mag. For the updated zeropoints we find a significant offset between the low-$z$ samples and high-$z$ samples, but no significant redshift dependent slope beyond z=0.1. 

For these combinations we also report the recovered dark energy equation of state $w$ relative to our replication of the original B14 analysis in Table \ref{tab:cosmoresults}. We note again that these $w$-differences are only for the subset of $\sim$700 that were calibrated for JLA. {\blue We also provide the same combination of changes to the analysis but relative to the SuperCal calibration \citep{Scolnic16} and SALT2 model trained on SuperCal offsets (T21). We find that the effect of changing both the SALT2 model and the zeropoints simultanously results in twice as large of a difference in recovered cosmology in comparison to only changing one component at a time (as did the original Pantheon). Relative to the original Pantheon, when accounting for the updated zeropoints and SALT2 model simultaneously we find $\Delta w=+0.07$ (0.064--0.009) on the subset of SNe that have been calibrated by both JLA and Fragilisitc.}

\subsection{Systematics Impact on Cosmological Inference}
\label{sec:syscosmo}

We analyze constraints on cosmological parameters before and after including the systematic covariance matrix and we find that the covariance inflates the posterior uncertainty on $w$ ($\sigma w_{\rm sys}$) by 0.013 after combining with Planck \cite{planck}. The breakdown for each component of calibration systematics analyzed in this work can be found in Table \ref{tab:sysdescription}. We note that this methodology has accounted for correlated SALT2 training and photometry zeropoints simultaneously, as described in Section~\ref{sec:sys}. This systematic uncertainty also accounts for novel conservative sources of systematics.

\section{Discussion}
\label{sec:discussion}

In previous analyses such as \cite{Scolnic18}, \citealt{Jones19}, \cite{Brout18b}, the systematics due to SALT2 retraining were decoupled from the shifts in calibration zeropoints.  With our formalism, we have simultaneously retrained SALT2 and fit light curves with our updated calibration. {\blue This was not done in previous analyses like \cite{Scolnic18} or \cite{Brout18b} because the SALT2 training code was neither easily available nor modifiable.}  As seen in Sec.~\ref{sec:cosmology}, the net change to $w$ for combined re-fitting with new zeropoints and re-training the SALT2 model is 0.06 relative to the Joint Light Curve Analysis and 0.07 relative to the original Pantheon analysis. While the calibration solution in this work largely agrees with that found in SuperCal, the large difference in $w$ is the result of the original Pantheon analysis not retraining the SALT2 model on their calibration and the difference in trained SALT2 models can be traced largely to the 1.5\% update in the CALSPEC modeling. For this reason in the Pantheon+ cosmological analysis we have tripled the associated systematic uncertainty in CALSPEC modeling.

The systematic uncertainty due to the propagation of uncertainties in our joint solution presented in Sec.~\ref{sec:syscosmo} is 0.013 in $w$.  As the reported shift in $w$ relative to previous analyses is significant compared to the typical statistical precision on $w$ and the reported systematic uncertainty, it is important to understand what drives this sensitivity. We can see in Table~\ref{tab:cosmoresults} that the re-calibration and re-training cause roughly equal changes to $w$.  

The net contribution to the systematic uncertainty found here is small in comparison to the older methodology utilized in the original Pantheon which treated SALT2 training and filter zeropoints separately as well as computed systematic uncertainties for filter zeropoints in an uncorrelated manner; individual linear perturbations. Additionally the original Pantheon adopted an additional uncertainty due to the fact that they did not retrain the SALT2 model for their updated calibration (SuperCal).

{\blue For this analysis, we followed the systematic-covariance matrix established in \cite{Conley2010}.  We both retrain the model and refit the light-curves using the updated Fragilistic calibration solution, and propagate calibration uncertainties through to cosmology.  As seen in \cite{binningissinning}, some systematics (for example SN intrinsic variations) can be `self-calibrated' down in size with simply larger datasets, whereas calibration related systematics benefit less from self-calibration and thus effort to reduce them with improved calibration measurements and accurate accounting of their uncertainties is important.}

We do find significant sensitivity to the inclusion of the observed frame $U$ band measurements in SALT2 training. As PS1 only covers $griz$, we are unable to re-calibrate the observed frame $U$ band, which is included for surveys CfA3S and CfA3K and CfA4.  While the original Pantheon chose not to include observed frame $U$ band in the fitting of light-curves, it was still used in the training of the SALT2 model.  This is not completely necessary to determine the model for rest-frame $U$ {\blue as the rest-frame $U$ is redshifted to higher observer-frame wavelengths at high redshift and the model can be calibrated via optical band measurements of higher-$z$ SNe in the training. We test this in further detail in Appendix~\ref{sec:nou}.}

While writing this paper, a new training code SaltShaker as part of SALT3 - \cite{Kenworthy21} was developed.  This is improved compared to the SALT2 model due to improved estimation of uncertainties, better separation of color and light-curve stretch, and a training sample size over $2\times$ larger.  Work is ongoing to implement it in standard SN Ia cosmology frameworks, but the sensitivity to the $U$ band discussed above motivate the adoption of surfaces with a longer wavelength range calibrated by higher redshift data. Until this happens we advocate for the removal of observer frame U which corresponds to $z>0.8$ which for the Pantheon+ sample is largely SNLS.

\begin{table*}
\centering
\begin{tabular}{lrrr}
\toprule
    Calibration & SALT2 Surface Only $w-w_{\rm B14}$ & LC Zpt Only $w-w_{\rm B14 }$& SALT2 Surface \& LC Zpt $w-w_{\rm B14}$ \\
\hline
      SuperCal &  0.033 & -0.009 & 0.007 \\
      Fragilistic &  0.035 & 0.025  & 0.064 \\
\end{tabular}
\caption{The changes to the recovered value of $w$ due to changes described in this analysis when combining with CMB data. This is only run on the subset of CfA,CSP,SDSS,SNLS that was calibrated in the Join Light Curve Analysis.}
\label{tab:cosmoresults}
\end{table*}

In Section~\ref{sec:cosmology} we gave the impact on $w$ resulting from the changes to the calibration, but we are also able to determine the impact on $H_0$.  We find the impact from all sources of calibration on the level of $0.2$~km/s/Mpc, subdominant relative to the total statistical uncertainty on $H_0$ of $\sim1$~km/s/Mpc in the companion SH0ES paper \citep{riess22}. This is due to the use of SNe from the same surveys in both the second and third rung of the distance ladder and from the use of many different photometric systems for both the calibrators and overlapping in redshift at low-$z$.  This effect is quantified in \cite{Brownsberger21}.  While \cite{Brownsberger21}  only look at the impact of single, gray offsets per survey, they find a similar $0.2$~km/s/Mpc limit in the uncertainties when excluding prior information on the survey zeropoints. While our grey offsets are much smaller than that of the worst case scenario of \cite{Brownsberger21}, because we include both color dependent correlations and retrained SALT2 (as \cite{Brownsberger21} did not) we find that the sensitivity to cosmological parameters $w$ and $\Omega_{M}$ is even stronger than that given in \cite{Brownsberger21}.
 
 \cite{Currie20} studied two effects that we did not: variations across the focal plane for each survey as well as the impact of filter shifts {\red (as did \citealt{Betoule2014})}. \cite{Currie20} finds while the focal plane variations can improve scatter, the net changes in distance are small and are therefore not included here.  The reason we didn't include filter shifts is because of the complexity in $2\times$ the degrees of freedom in our simultaneous fit {\blue and introduction of non-linearity in the computation of the likelihood.}  Despite this, we took care to examine filters that had large differences in their slopes for observed and synthetic {\red and still varied the uncertainty on each filter curve in the SALT2 training systematics as shown in Table \ref{tab:wavunc} albeit without correlations with the zeropoints.}

The other way to improve this analysis is with more calibration data.  While future surveys (e.g. Vera Rubin LSST, Nancy Grace Roman Space Telescope) can move away from this cross-calibration method by using a single photometric system, this is currently not feasible for measuring $H_0$.  
While we are fortunate in that new surveys with many more stars and better calibration are coming (such as DES, ZTF, LSST and Roman), any improvement for $H_0$ will be relatively small due to the necessity of using SNe in older surveys {\blue that have measured the rare nearby SNe required for Cepheid observations over the last 20 years. }

\section{Conclusion}

In this analysis, we compiled the information to calibrate 25 different photometric systems used for the upcoming Pantheon+ and SH0ES analyses.  We measure calibration zeropoints for a total of 105 filters.  We find relatively good agreement with SuperCal, though add on an additional 8 photometric systems and account for updated CALSPEC modeling and CSP and CfA filter throughputs.  

We derive a full covariance matrix between the filter zeropoints; the first of its kind. We then retrain the SALT2 model based on our fitted zeropoints and use the covariance matrix to create 9 realizations based on correlated perturbed calibration offsets.  

{\blue We find a change in $w$ compared to the original Pantheon of 0.07 and a 30\% reduction in contribution to systematic calibration uncertainty with our new method.}

\newpage

\newpage

\newpage

\bibliography{paper}
\bibliographystyle{aasjournal}
\section{Acknowledgements}
 We thank Rick Kessler for his ever-useful SNANA package and his work to incorporate SALT2 training in the SNANA framework.
 DB acknowledges support for this work was provided by NASA through the NASA Hubble Fellowship grant HST-HF2-51430.001 awarded by the Space Telescope Science Institute, which is operated by Association of Universities for Research in Astronomy, Inc., for NASA, under contract NAS5-26555. DS is supported by DOE grant DE-SC0010007, DE-SC0021962 and the David and Lucile Packard Foundation. DS is supported in part by the National Aeronautics and Space Administration (NASA) under Contract No.~NNG17PX03C issued through the Roman Science Investigation Teams Programme. GT is supported by an Australian Government Research Training Program (RTP) Scholarship. This work was completed in part with resources provided by the University of Chicago’s Research Computing Center. Simulations, light-curve fitting, BBC, and cosmology pipeline managed by \texttt{PIPPIN} \citep{Pippin}. Contours and parameter constraints are generated using the \textsc{ChainConsumer} package \citep{Hinton16}. Plots generated with Matplotlib \citep{matplotlib}. Usage of astropy \citep{astropy}, SciPy \citep{scipy}, and NumPy \citep{numpy}.
 
 Brout thanks his spouse Isabella and their future daughter for their support as the due date is rapidly approaching!

\appendix

\section{SALT2 Retrained for Systematic Uncertainty}

\begin{figure*}
        \centering 
	    \includegraphics[width=.48\textwidth]{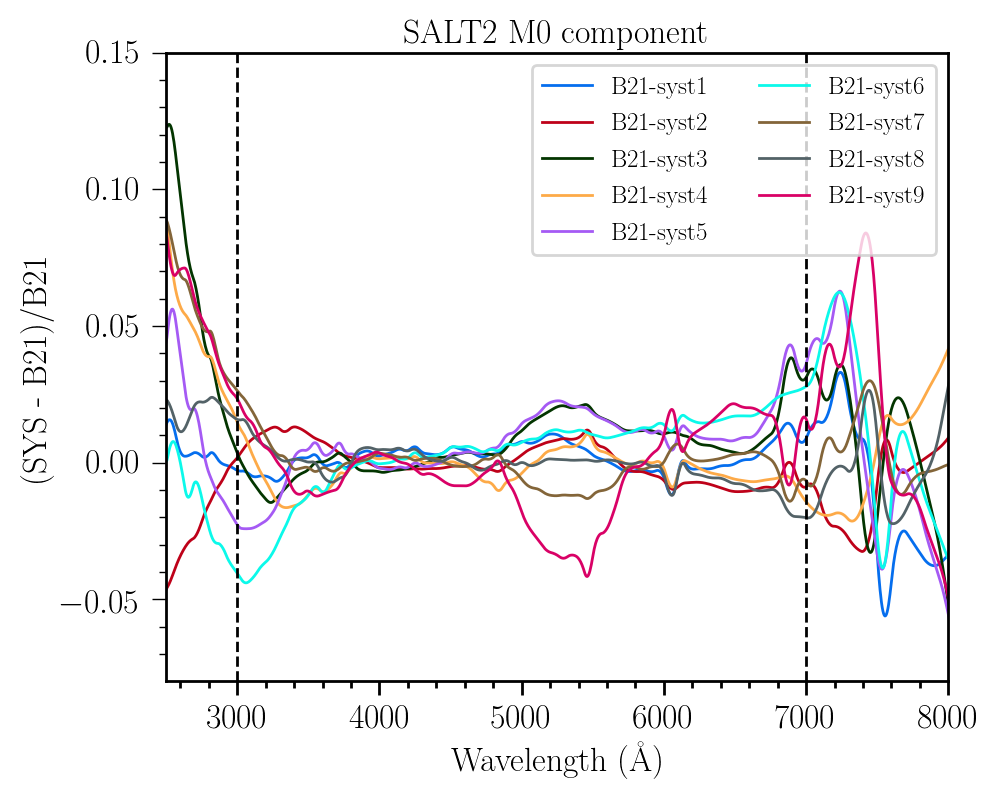} 
        	    \includegraphics[width=.48\textwidth]{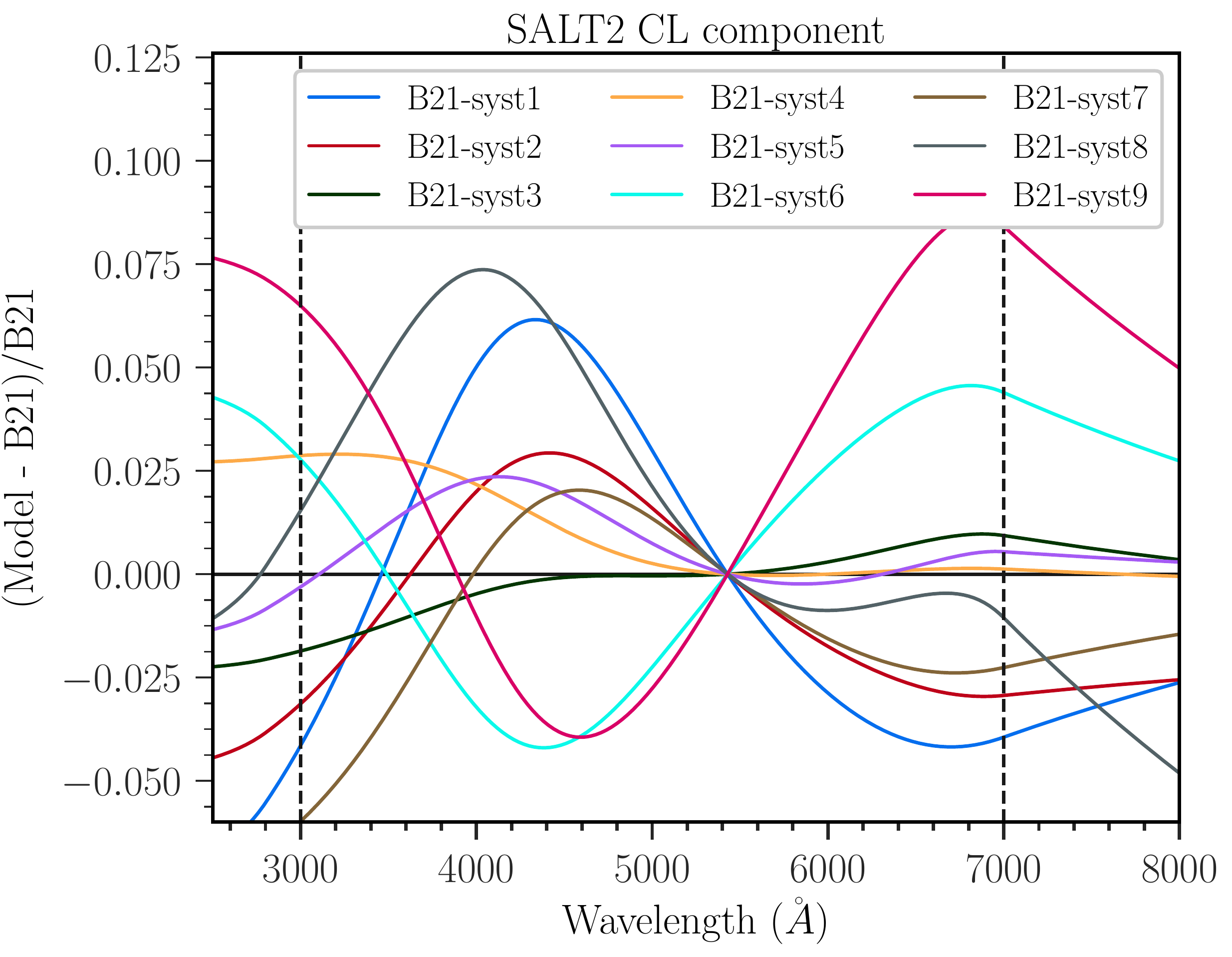} 
\caption{Same as Figure~\ref{fig:saltsurfb14}, but showing the impact of the 9 different systematic perturbations with correlated zeropoint offsets.}
        \label{fig:saltsurfsys} 
    \end{figure*}

\begin{figure}
        \centering 
    
        \includegraphics[width=.48\textwidth]{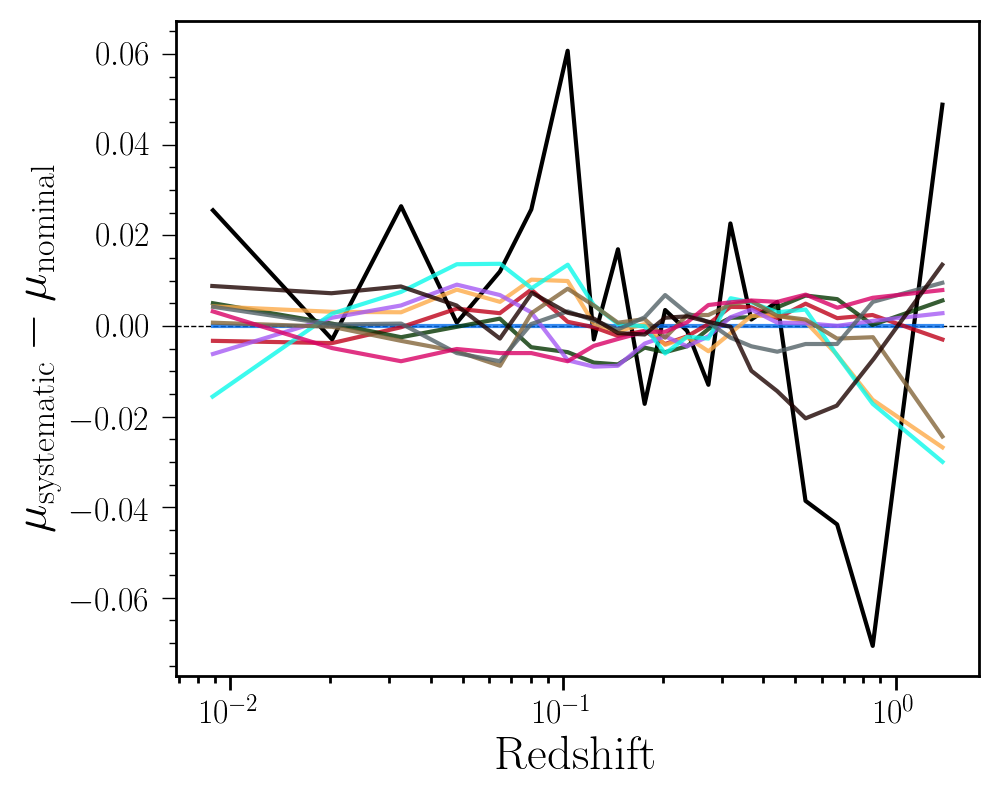} 
        \caption{Binned Hubble diagram residuals relative to the nominal Fragilistic calibration and B21 SALT2 model. Each curve represents a different set of correlated zeropoint offsets propagated through SALT2 training, light-curve fitting, and standardization (Eq.~\ref{Eq:tripp}).}
        \label{fig:salt2retrainmudif} 
    \end{figure}

Similar to \cite{Betoule2014}, we retrain the SALT2 model for several systematic variants to our analysis. We utilize the Fragilistic zeropoint covariance matrix to produce realizations of calibration solutions with correlated zeropoints. For each realization we also allow for uncorrelated variation in the effective mean wavelengths of each filter. The uncertainties used are documented in Table~\ref{tab:wavunc}. We note that based on Table~\ref{tab:sensitivity}, these are not the dominant source of uncertainty in distances in comparison to the filter zeropoints. The resulting 9 SALT2 M0 components and color laws CL are shown in Fig.~\ref{fig:saltsurfsys}. When using the 9 SALT2 models and sets of calibration offsets simultaneously we obtain the 9 Hubble diagrams where the residuals to the nominal fit are shown in Fig.~\ref{fig:salt2retrainmudif}.

\begin{table}
\centering
\begin{tabular}{l|rr}
\toprule
Sample & Filters & Wave. Uncertainty (\AA) \\
\hline
Calan/Tololo  & $U,B,V,R,I$ &  20,20,20,20,20 \\
CfA1  & $U,B,V,R,I$ &  20,20,20,20,20 \\
CfA2  & $U,B,V,R,I$ & 20,20,20,20,20 \\
CfA3-Keplercam  & $U,B,V,r,i$ & 20,10,10,10,10\\
CfA3-4Shooter  &$U,B,V,r,i$ &  20,10,10,10,10  \\
CSP  &$u,B,V,g,r,i$ &  8,7,3,8,4,2  \\
SDSS  &  $u,g,r,i,z$ & 6,6,6,6,6\\
SNLS  &  $g,r,i,z$ &  3,10,10,6\\
\end{tabular}
\caption{List of the photometric data samples in SALT2 training for which effective mean wavelengths were varied. Gaussian $1\sigma$ uncertainties used for the variation are reported.
}
\label{tab:wavunc}
\end{table}

\section{SALT2 Reliance on Rest-frame $U$ Band}
\label{sec:nou}
\begin{figure*}
        \centering 
	    \includegraphics[width=.46\textwidth]{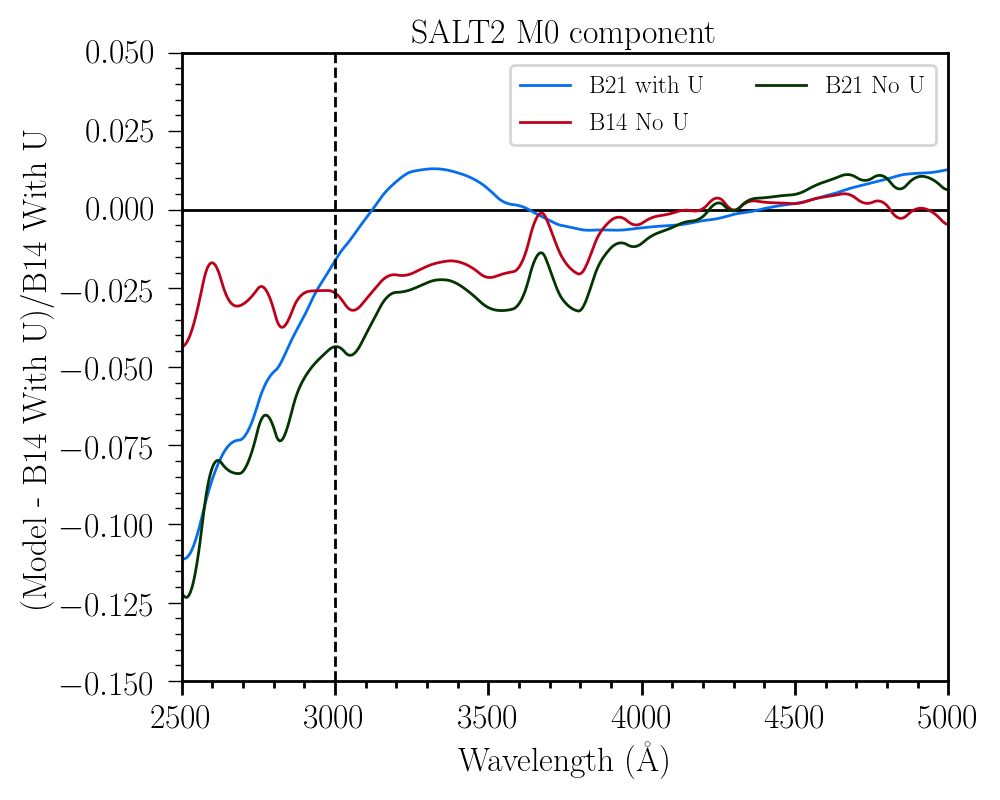} 	    \includegraphics[width=.46\textwidth]{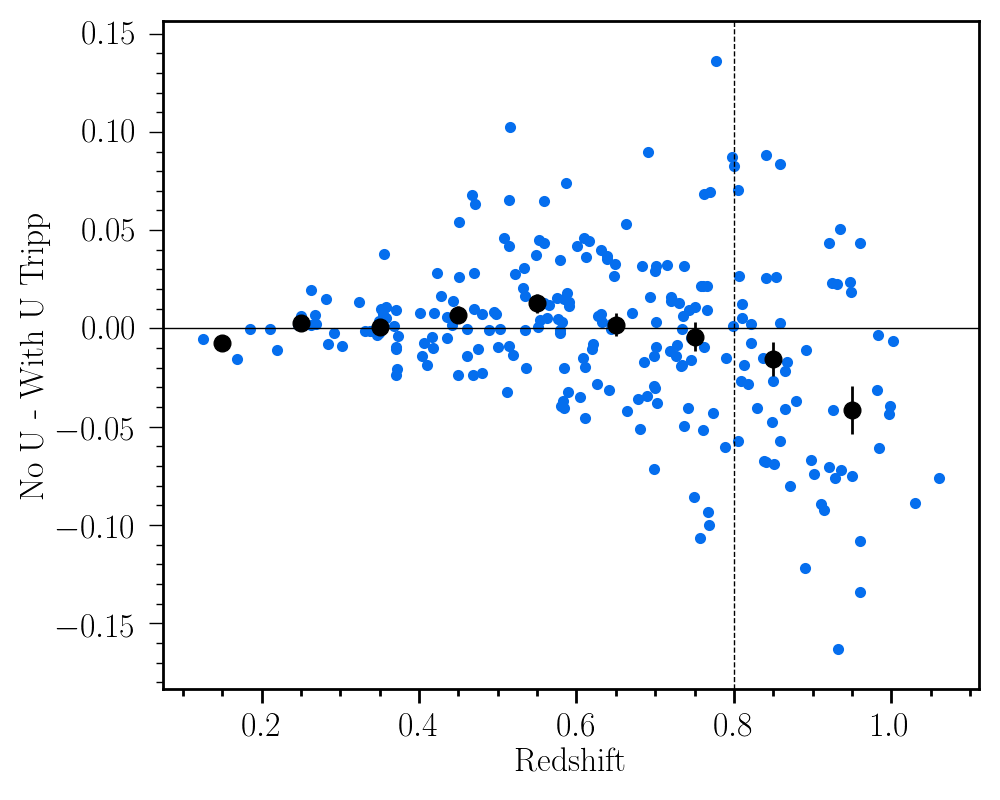} 
        \caption{{\blue [Left] Fractional differences in the M0 component of the SALT2 model relative to the B14 model with and without rest-frame $U$ band included in the training. [Right] Differences in inferred distances assuming the same Tripp standardization coefficients $\alpha=0.15$ and $\beta=3.1$ between fits with and without rest-frame $U$ band in the SALT2 training. Differences are largest beginning at $z=0.8$. A vertical line is placed at $z=0.8$ where we place a quality cut in Pantheon+ sample \citep{scolnic22} in order to be insensitive to this potential systematic.}}
        \label{fig:nou} 
    \end{figure*}

 {\blue We test the impact of including/excluding observer-frame $U$ band photometry in the SALT2 training sample and compute distances for the SNLS dataset (Fig.~\ref{fig:nou}).} For the original JLA or new calibration, the impact of not including $U$, which is notoriously difficult to calibrate, is very large, and causes a shift in $w$ on the order of $\sim0.05$, with the highest effect at high-$z$.  Therefore, because of the difficulty of calibrating $U$ but likely necessity for smoothness in the training, we used $U$ in the training but in S21 do not include SNe in surveys at redshifts that are particularly sensitive to rest-frame $U$. {\blue We place a cut on $z>0.8$ for optical $griz$ ground based surveys which affects 2 SNe from DES-SN3YR and 57 SNe from SNLS.}
    
\section{Description of the calibration of each system}

Here we present the filled out templates for each photometric system cross calibration in this analysis.

\begin{table*}
\label{table:information}
\caption{\large CfA1}
\begin{tabular}{lr}
\toprule
    Question & Answer \\
\hline

Where are the filters published? & \cite{CfA1} \\
Where is the site for the filters? & -\\
Does filter have atmospheric term?	& Yes \\
Where are the stars published? & \cite{Currie20} \\
Where is the site for the stars? & \href{https://arxiv.org/pdf/2007.02458.pdf }{Here} \\
What system are the stars in? & Landolt \\
What is calibration based on? & BD17 \\
\end{tabular}
\end{table*}

\begin{table*}
\caption{\large CfA2}
\begin{tabular}{lr}
\toprule
    Question & Answer \\
\hline

Where are the filters published? & \cite{CfA2} \\
Where is the site for the filters? & \href{https://iopscience.iop.org/article/10.1086/497989/fulltext/204512.tables.html}{Here} \\
Does filter have atmospheric term?	& Yes \\
Where are the stars published? & \cite{cfa2stars} \\
Where is the site for the stars? & \href{https://iopscience.iop.org/article/10.1086/497989/fulltext/204512.tables.html}{Here} \\
What system are the stars in? & Standard \\
What is calibration based on? & Landolt,BD17 \\
\end{tabular}
\end{table*}

\begin{table*}
\label{table:information}
\caption{\large CfA3-Keplercam}
\begin{tabular}{lr}
\toprule
    Question & Answer \\
\hline

Where are the filters published? & \cite{CfA3-Keplercam} \\
Where is the site for the filters? & https://www.cfa.harvard.edu/supernova/CfA3/ \\
Does filter have atmospheric term?	& No \\
Where are the stars published? & \cite{cfa3keplerstars} \\
Where is the site for the stars? & https://www.cfa.harvard.edu/supernova/CfA3/ \\
What system are the stars in? & Standard,Natural \\
What is calibration based on? & Landolt,BD17 \\
What is transformation for that calibration? & ~\\
~& $(u-b)/(U-B) = 1.0279 \pm 0.0069$ \\
~ & $(b-v)/(B-V) = 0.9212 \pm 0.0029$ \\
~ & $(v-V)/(B-V) = 0.0185 \pm 0.0023$ \\
~ & $(v-r)/(V-r'') = 1.0508 \pm 0.0029$ \\
~ & $(v-i)/(V-i'') = 1.0185 \pm 0.0029$ \\
\end{tabular}
\end{table*}

\begin{table*}
\caption{\large CfA3-4Shooter}
\label{table:information}
\begin{tabular}{lr}
\toprule
    Question & Answer \\
\hline

Where are the filters published? & \cite{CfA3-4Shooter} \\
Where is the site for the filters? & https://iopscience.iop.org/article/10.1086/497989/fulltext/204512.tables.html \\
Does filter have atmospheric term?	& Yes \\
Where are the stars published? & \cite{cfa34shooterstars} \\
Where is the site for the stars? & https://www.cfa.harvard.edu/supernova/CfA3/ \\
What system are the stars in? & Standard and Natural \\
What is calibration based on? & Landolt,BD17 \\
What is transformation for that calibration? & ~\\
~& $(u-b)/(U-B) = 0.9912 \pm 0.0078$ \\
~ & $(b-v)/(B-V) = 0.8928 \pm 0.0019$ \\
~ & $(v-V)/(B-V) = 0.0336 \pm 0.002$ \\
~ & $(v-r)/(V-R) = 1.0855 \pm 0.0058$ \\
~ & $(v-i)/(V-I) = 1.0166 \pm 0.0067$ \\
\end{tabular}
\end{table*}

\begin{table*}
\label{table:information}
\caption{\large CfA4 p1}
\begin{tabular}{lr}
\toprule
    Question & Answer \\
\hline

Where are the filters published? & \cite{CfA4_p1} \\
Where is the site for the filters? & https://www.cfa.harvard.edu/supernova/CfA4/ \\
Does filter have atmospheric term?	& Yes \\
Where are the stars published? & \cite{cfa4p1stars} \\
Where is the site for the stars? & https://www.cfa.harvard.edu/supernova/CfA4/ \\
What system are the stars in? & Standard and Natural \\
What is calibration based on? & Landolt,BD17 \\
What is transformation for that calibration? & ~\\
~& $(u-b)/(U-B) = 0.9981 \pm 0.0209$ \\
~ & $(u-b)/(u'-B) = 0.8928 \pm 0.0057$ \\
~ & $(b-v)/(B-V) = 0.0336 \pm 0.0026$ \\
~ & $(v-V)/(B-V) = 1.0855 \pm 0.0018$ \\
~ & $(v-r)/(V-r') = 1.0166 \pm 0.0028$ \\
~ & $(v-i)/(V-i') = 1.0166 \pm 0.0016$ \\
\end{tabular}
\end{table*}

\begin{table*}
\label{table:information}
\caption{\large CSP DR3}
\begin{tabular}{lr}
\toprule
    Question & Answer \\
\hline

Where are the filters published? & \cite{CSPDR3} \\
Where is the site for the filters? & \url{https://csp.obs.carnegiescience.edu/data/filters}  \\
Does filter have atmospheric term?	& Yes \\
Where are the stars published? & \cite{cspdr3stars} \\
Where is the site for the stars? & https://iopscience.iop.org/1538-3881/160/6/289/suppdata/ajabc431t5\_mrt.txt \\
What system are the stars in? & Natural \\
What is calibration based on? & Landolt,Smith,BD17 \\
What is transformation for that calibration? & ~\\
~& $u = u'-cu*(u'-g') = 0.046 \pm 0.017$ \\
~ & $g = g'-cg*(g'-r') = -0.014 \pm 0.011$ \\
~ & $r = r'-cr*(r'-i') = -0.016 \pm 0.015$ \\
~ & $i = i'-ci*(r'-i') = -0.002 \pm 0.015$ \\
~ & $B = B'-cB*(B'-V') = 0.061 \pm 0.012$ \\
~ & $V = V'-cV*(V'-i') = -0.058  \pm 0.011$ \\
\end{tabular}
\end{table*}

       \begin{figure}
        \centering 
	    \includegraphics[width=.6\textwidth]{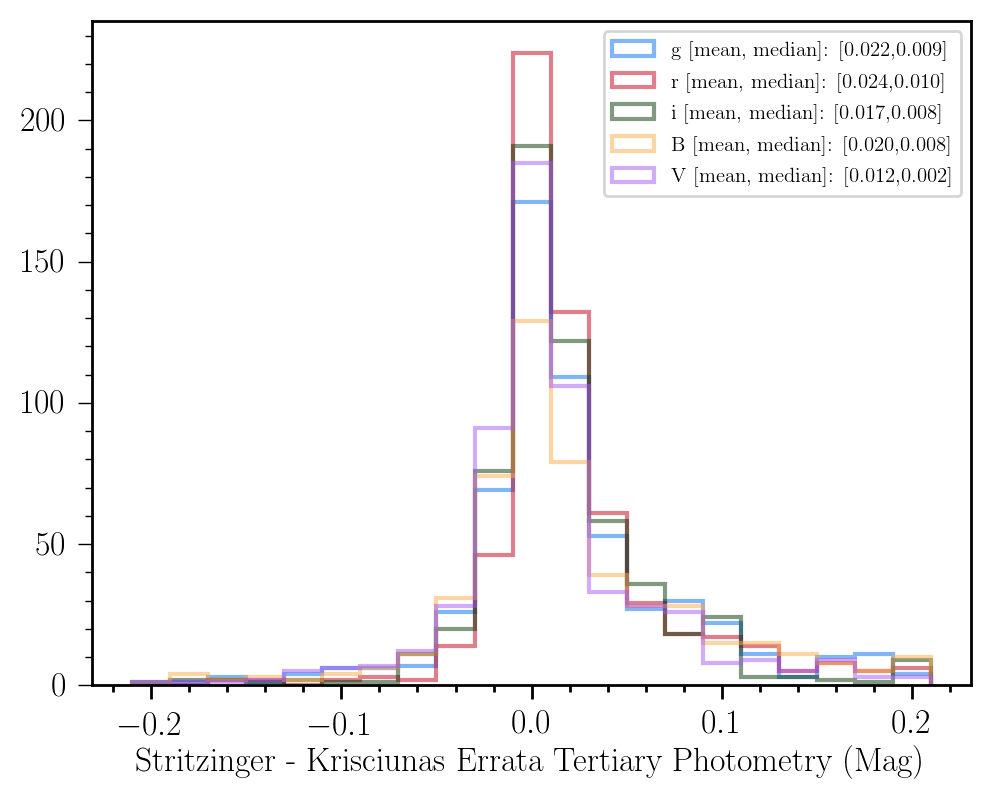} 
        \caption{Difference in stellar mags between the tertiary catalogs of \cite{Stritzinger2010} and erratum \cite{krisciunas20}.}
        \label{fig:cspudpate} 
    \end{figure}

\begin{table*}
\label{table:information}
\caption{\large LOSS1 SN Survey}
\begin{tabular}{lr}
\toprule
    Question & Answer \\
\hline

Where are the filters published? & \cite{KAIT-LOSS} \\
Where is the site for the filters? & Private Communication \\
Does filter have atmospheric term?	& Yes \\
Where are the stars published? & Private Communication \\
Where is the site for the stars? & Private Communication\\
What system are the stars in? & AB \\
What is calibration based on? & PS1 \\
What is transformation for that calibration? & ~\\
~& $b = B+CB*(B-V) = -0.095$ \\
~ & $v = V+CV*(B-V) = 0.027$ \\
~ & $r = R+CR*(V-R) = -0.181$ \\
~ & $i = I+CI*(V-I) = -0.071$ \\
\end{tabular}
\end{table*}

\begin{table*}
\label{table:information}
\caption{\large LOSS2 SN Survey}
\begin{tabular}{lr}
\toprule
    Question & Answer \\
\hline

Where are the filters published? & \cite{Ganeshalingam10} \\
Where is the site for the filters? & Private Communication \\
Does filter have atmospheric term?	& Yes \\
Where are the stars published? & \cite{Ganeshalingam10}  \\
Where is the site for the stars? & \href{https://cfn-live-content-bucket-iop-org.s3.amazonaws.com/journals/0067-0049/190/2/418/1/apjs341025t7_mrt.txt?AWSAccessKeyId=AKIAYDKQL6LTV7YY2HIK&Expires=1639449948&Signature=g%2B%2BlR%2Brz3qogjr%2FOiE5zsWjSkiU%3D}{Here} \\
What system are the stars in? & Landolt \\
What is calibration based on? & BD17 \\
What is transformation for that calibration? & ~\\
~& $b = B+CB*(B-V) = -0.095$ \\
~ & $v = V+CV*(B-V) = 0.027$ \\
~ & $r = R+CR*(V-R) = -0.181$ \\
~ & $i = I+CI*(V-I) = -0.071$ \\
\end{tabular}
\end{table*}

\begin{table*}
\label{table:information}
\caption{\large SOUSA SN Survey}
\begin{tabular}{lr}
\toprule
    Question & Answer \\
\hline

Where are the filters published? & \cite{brown14} \\
Where is the site for the filters? & \href{https://ui.adsabs.harvard.edu/abs/2011AIPC.1358..373B/abstract}{Here} \\
Does filter have atmospheric term?	& No \\
Where are the stars published? & Private communication \\
Where is the site for the stars? & \href{http://vizier.cfa.harvard.edu/viz-bin/VizieR-3?-source=J/ApJS/190/418/table7}{Here} \\
What system are the stars in? & (Natural/Landolt/AB) \\
What is calibration based on? & Specific standard (e.g. BD17) \\
What is transformation for that calibration? & ~\\
~& $U-V = 0.087+0.8926(u-v)+0.0274(u-v)^2$ \\
~ & $B-V = 0.0148+1.0184(b-v)$ \\
~ & $B = b+0.0173+0.0187(u-b)+0.013(u-b)^2-0.0108(u-b)^3$\\
~ & $-0.0058(u-b)^4+0.0026(u-b)^5$ \\
~ & $V = v+0.0006-0.0113(b-v)+0.0097(b-v)^2-0.0036(b-v)^3$ \\
\end{tabular}
\end{table*}

\begin{table*}
\label{table:information}
\caption{\large PS1 SN Survey}
\begin{tabular}{lr}
\toprule
    Question & Answer \\
\hline

Where are the filters published? & \cite{PS1} \\
Where is the site for the filters? & \href{http://vizier.cfa.harvard.edu/viz-bin/VizieR?-source=J/ApJ/750/99}{http://vizier.cfa.harvard.edu/viz-bin/VizieR?-source=J/ApJ/750/99} \\
Does filter have atmospheric term?	& Yes \\
Where are the stars published? & \cite{Scolnic18}\\
Where is the site for the stars? & Private communication \\
What system are the stars in? & AB \\
What is calibration based on? & Multiple standards \\
\end{tabular}
\end{table*}

\begin{table*}
\label{table:information}
\caption{\large Foundation SN Survey}
\begin{tabular}{lr}
\toprule
    Question & Answer \\
\hline

Where are the filters published? & \cite{Foundation} \\
Where is the site for the filters? & \href{http://vizier.cfa.harvard.edu/viz-bin/VizieR?-source=J/ApJ/750/99}{http://vizier.cfa.harvard.edu/viz-bin/VizieR?-source=J/ApJ/750/99} \\
Does filter have atmospheric term?	& Yes \\
Where are the stars published? & Private Communication \\
Where is the site for the stars? & Private Communication \\
What system are the stars in? & AB \\
What is calibration based on? & Multiple standards \\
\end{tabular}
\end{table*}

\begin{table*}
\label{table:information}
\caption{\large DES SN 3-Year}
\begin{tabular}{lr}
\toprule
    Question & Answer \\
\hline

Where are the filters published? & \cite{DES} \\
Where is the site for the filters? & - \\
Does filter have atmospheric term?	& Yes \\
Where are the stars published? & \cite{fgcm} \\
Where is the site for the stars? & Private Communication \\
What system are the stars in? & (Natural/Landolt/AB) \\
What is calibration based on? & C26202 \\
\end{tabular}
\end{table*}

\begin{table*}
\label{table:information}
\caption{\large SDSS SN Survey}
\begin{tabular}{lr}
\toprule
    Question & Answer \\
\hline

Where are the filters published? & \cite{SDSS} \\
Where is the site for the filters? & \href{https://iopscience.iop.org/article/10.1088/0004-6256/139/4/1628/pdf}{https://iopscience.iop.org/article/10.1088/0004-6256/139/4/1628/pdf} \\
Does filter have atmospheric term?	& Yes \\
Where are the stars published? & \cite{sdssstars} \\
Where is the site for the stars? & \href{http://cdsarc.u-strasbg.fr/cgi-bin/qcat?J/A+A/552/A124}{http://cdsarc.u-strasbg.fr/cgi-bin/qcat?J/A+A/552/A124} \\
What system are the stars in? & AB \\
What is calibration based on? & Multiple standards \\

\end{tabular}
\end{table*}

\begin{table*}
\label{table:information}
\caption{\large SNLS SN Survey}
\begin{tabular}{lr}
\toprule
    Question & Answer \\
\hline

Where are the filters published? & \cite{SNLS} \\
Where is the site for the filters? & \href{http://cdsarc.u-strasbg.fr/cgi-bin/qcat?J/A+A/552/A124}{http://cdsarc.u-strasbg.fr/cgi-bin/qcat?J/A+A/552/A124} \\
Does filter have atmospheric term?	& Yes \\
Where are the stars published? & \cite{sdssstars} \\
Where is the site for the stars? & \href{http://cdsarc.u-strasbg.fr/cgi-bin/qcat?J/A+A/552/A124}{http://cdsarc.u-strasbg.fr/cgi-bin/qcat?J/A+A/552/A124} \\
What system are the stars in? & AB \\
What is calibration based on? & Multiple standards \\

\end{tabular}
\end{table*}

\section{Re-assessment of the original calibrations for PS1, SDSS, SNLS, DES due to changes in CALSPEC}

\begin{table}
\centering
\begin{tabular}{lc}
\toprule
    Survey & Mag Offset g,r,i,z \\
\hline
      PS1/Foundation & 0.011,0.012,0.008,0.008  \\
      SDSS & 0.005,-0.006,-0.007,-0.007  \\
      SNLS & 0.005,-0.006,-0.007,-0.007 \\
       DES & 0.011,0.011,0.008,0.004 \\
\label{tab:recal}
\end{tabular}
\caption{AB recalibration for AB surveys that we calibrated on old CALSPEC models. These offsets are applied prior to performing cross-calibration.}
\end{table}

\end{document}